\documentclass{article}

\usepackage{PRIMEarxiv}

\usepackage[utf8]{inputenc} 
\usepackage[T1]{fontenc}    
\usepackage{hyperref}       
\usepackage{url}            
\usepackage{booktabs}       
\usepackage{amsfonts}       
\usepackage{nicefrac}       
\usepackage{microtype}      
\usepackage{lipsum}
\usepackage{fancyhdr}       
\usepackage{graphicx}       
\graphicspath{{media/}}     

\usepackage{amsmath,mathabx}
\usepackage{bm}
\usepackage{cite}
\usepackage{algorithm}
\usepackage{algpseudocode}
\usepackage{enumitem}
\usepackage{threeparttable}
\usepackage{setspace}
\usepackage[toc,page]{appendix}

\let\appendixpagenameorig\appendixpagename
\renewcommand{\appendixpagename}{\Large\appendixpagenameorig}

\title{A Dirichlet Process Mixture Model for Directional-Linear Data

}

\author{
  Tong Zou \\
  Department of Statistics \\
  University of California, Irvine \\
  Irvine\\
  \texttt{zout3@uci.edu} \\
   \And
  Hal Stern \\
  Department of Statistics \\
  University of California, Irvine \\
  Irvine\\
  \texttt{sternh@uci.edu} \\
}

\begin{document}
\maketitle

\begin{abstract}

Directional data require specialized probability models because of the non-Euclidean and periodic nature of their domain. When a directional variable is observed jointly with linear variables, modeling their dependence adds an additional layer of complexity. This paper introduces a novel Bayesian nonparametric approach for directional-linear data based on the Dirichlet process. We first extend the projected normal distribution to model the joint distribution of linear variables and a directional variable with arbitrary dimension as a projection of a higher-dimensional augmented multivariate normal distribution (MVN). We call the new distribution the semi-projected normal distribution (SPN); it possesses properties similar to the MVN. The SPN is then used as the mixture distribution in a Dirichlet process model to obtain a more flexible class of models for directional-linear data. We propose a normal conditional inverse-Wishart distribution as part of the prior distribution to address an identifiability issue inherited from the projected normal and preserve conjugacy with the SPN distribution. A Gibbs sampling algorithm is provided for posterior inference. Experiments on synthetic data and the Berkeley image database show superior performance of the Dirichlet process SPN mixture model (DPSPN) in clustering compared to other directional-linear models. We also build a hierarchical Dirichlet process model with the SPN to develop a likelihood ratio approach to bloodstain pattern analysis using the DPSPN model for density estimation to estimate the likelihood of a given pattern from a set of training data.

\end{abstract}


\section{Introduction}
\label{intro}
Directional statistics is the subdiscipline of statistics used to study directional observations that can be represented as unit vectors in Euclidean space. The sample space of a directional variable denoted by a unit vector in $\mathbb{R}^p$ is the surface of an $(p-1)$ dimensional unit hypersphere $\mathbb{S}^{p-1}$. The most common directional observations are circular data and spherical data in the cases that $p=2$ and $p=3$. Directional observations arise in many scientific fields and applications, including studies of wind directions \cite{jammalamadaka2006effect, nunez2015bayesian}, motion planning for robots \cite{kucner2017enabling, palmieri2017kinodynamic}, and image analysis \cite{hasnat2014unsupervised, roy2016swgmm}. Due to the non-Euclidean periodic property of the domain of directional observations, the analysis of such data requires specialized statistical models. Examples of parametric models include the von Mises-Fisher family \cite{watson1982distributions} and variants of the normal distribution (e.g., the projected normal \cite{mardia1975statistics,wang2013directional, hernandez2017general} and wrapped normal \cite{collett1981discriminating}). See \cite{mardia2000directional} and \cite{pewsey2021recent} for a more comprehensive review. A number of recent studies focus on more flexible modeling of directional data using mixtures of distributions, including mixtures of the normal variants \cite{nunez2015bayesian, wang2014modeling,rodriguez2020bayesian}, mixtures of von Mises distributions \cite{carta2008statistical} and sums of trigonometric functions \cite{fernandez2004circular}.

Directional data are often observed together with linear variables that take values on the real line. For example, in the study of meteorology, wind directions may be collected along with other linear components like wind speed, temperature and humidity \cite{fernandez2007models}; and in image analysis, some color spaces adopt hue (a circular variable) and other linear measurements (e.g., chroma and lightness) to represent color information. Establishing the joint distribution of directional-linear data requires modeling the correlation of directional and linear components. This is not a straightforward task due to the complex manifold of the sample space. In the case of one circular variable and one linear variable, the sample space is the surface of a cylinder. A popular approach to modeling cylindrical data is to use a copula density that can marginalize to a circular distribution and a linear distribution \cite{johnson1978some, fernandez2007models,carta2008joint,soukissian2014probabilistic,zhang2018investigation}. \textit{Roy et al.} \cite{roy2017jclmm} developed mixtures of copula distributions to obtain a more flexible family of models. To the best of our knowledge, the copula models developed so far are all bivariate models for circular-linear data, and extending them into higher-dimensional space (e.g., by including a spherical variable or additional linear variables) is not a trivial problem. Efforts have been made to model the joint distribution of directional-linear data in higher dimensional space based on the multivariate normal distribution (MVN), which conveniently models the correlations among multiple variables. For example, by transforming one dimension of the MVN into a wrapped normal, the MVN distribution is capable of modeling the joint distribution of one circular variable and multiple linear variables. \textit{Roy et al.} \cite{roy2016swgmm} developed a mixture model based on this idea. \textit{Mastrantonio} \cite{mastrantonio2018joint} proposed using projected normal and skew-normal distributions to model the joint distribution of multiple circular variables and linear variables. All of the models mentioned above are limited to circular data and not applicable to directional variables in higher dimensions (e.g., a spherical variable).

In this study we propose a novel approach to modeling directional-linear data that can accommodate a directional variable with dimension $p > 2$. The basic idea is to use a MVN to derive the joint distribution of multiple linear variables and
one directional variable in arbitrary dimension, and then marginalize to a projected normal. Following similar nomenclature as in \cite{roy2016swgmm}, we call the resulting marginal distribution a semi-projected normal distribution (SPN). With appropriate covariance structure, the SPN can accommodate skewed and bimodal distributions for the directional component. In many real-world applications, data distributions can be multimodal and too complex for a specified parametric distribution. Nonparametric Bayesian approaches like Dirichlet process mixture models (DPMM) are often applied to address such cases. We define a DPMM based on the SPN to create a more flexible model for directional-linear data and implement a Markov chain Monte Carlo sampling algorithm to fit the model. The projected normal distribution requires a constraint on the covariance matrix to ensure the model is identifiable \cite{hernandez2017general, mastrantonio2018joint}, and our Bayesian approach requires a prior distribution for the covariance matrix that is subject to the same constraint. We developed a conditional inverse-Wishart distribution to accommodate the constraint and still take advantage of conjugacy to achieve efficient sampling.

The remainder of this paper is organized as follows. Section \ref{SPN} reviews the definition and properties of the projected normal distribution and introduces the SPN for directional-linear data. Section \ref{secdpmm} reviews the basic setting of the DPMM and then develops a DPMM based on the SPN to build a more flexible and robust model for directional-linear data. Section \ref{Experiments} applies our Dirichlet process SPN mixture model (DPSPN) to clustering of synthetic data and image segmentation and compares its performance with other state-of-the-art methods. Section \ref{BPA} develops a hierarchical DPSPN that is applied to density estimation for bloodstain pattern analysis. A summary discussion is provided in Section \ref{Discussion}.

\section{The semi-projected normal distribution}
\label{SPN}

In this section, we first review the projected normal distribution that can be used to model a directional variable with arbitrary dimension. Then we introduce the semi-projected normal distribution (SPN) as a generalization of the projected normal to model the joint distribution of directional-linear data.

\subsection{The projected normal distribution for directional data}

One approach to obtaining a distribution for directional data is by projecting a distribution defined on $\mathbb{R}^p$ onto the unit hypersphere $\mathbb{S}^{p-1}$. For $p\geqslant 2$, let the random vector $\bm{x} = (x_1,...,x_p)^T$ follow a $p$-variate normal distribution $\mathcal{N}_p(\bm{\mu},\bm{\Sigma})$ and define the directional variable $\bm{u} = (u_1,...,u_p)^T = r^{-1} \bm{x}$ where $r = \lVert \bm{x} \rVert = (\bm{x}^T\bm{x})^{\frac{1}{2}}$ is the radius. The marginal distribution of $\bm{u}$ is called the projected normal with parameters $\bm{\mu}$ and $\bm{\Sigma}$ and denoted by $\mathcal{PN}_p(\bm{\mu},\bm{\Sigma})$. It is defined on the unit hypersphere in $(p-1)$ dimensions $\mathcal{S}^{p-1}$. An alternative way to represent $\bm{u}$ is to use $(p-1)$ angular coordinates $\bm{\theta} = (\theta_1,...,\theta_{p-1})$ in the spherical system, where $\theta_1,...,\theta_{p-2}$ range over $[0,\pi]$ and $\theta_{p-1}$ ranges over $[0,2\pi)$. Since $\bm{\theta}$ and $\bm{u}$ represent the same direction, they are often used interchangeably in the literature. The vector $\bm{x}$ can be computed from $r$ and $\bm{\theta}$ via the following transformation:
\begin{align} \label{car2sph}
\begin{split}
    x_1 &= r\cos{\theta_1} \\
    x_2 &= r\sin{\theta_1}\cos{\theta_2} \\
    x_3 &= r\sin{\theta_1}\sin{\theta_2}\cos{\theta_3} \\
    ... \\
    x_{p-1} &= r\sin{\theta_1}...\sin{\theta_{p-2}}\cos{\theta_{p-1}} \\
    x_{p} &= r\sin{\theta_1}...\sin{\theta_{p-2}}\sin{\theta_{p-1}}
\end{split}
\end{align}
The joint density function of $r$ and $\bm{\theta}$ can be derived from the density of $\mathcal{N}_p(\bm{\mu},\bm{\Sigma})$ and the Jacobian matrix of the transformation (\ref{car2sph}):
\begin{align}
\begin{split}
    f(r,\bm{\theta}|\bm{\mu},\bm{\Sigma})
    &= f_{\bm{X}}(\bm{x}|\bm{\mu},\bm{\Sigma})|\frac{\partial\bm{x}}{\partial(r,\bm{\theta})}| \\
    &= |2\pi\bm{\Sigma}|^{-\frac{1}{2}}r^{p-1}\exp{\Bigl\{-\frac{1}{2}(r\bm{u}-\bm{\mu})^T\bm{\Sigma}^{-1}(r\bm{u}-\bm{\mu})\Bigl\}}\prod_{j=1}^{p-2}(\sin{\theta_j})^{p-1-j} \label{rthetajoint}
\end{split}
\end{align}
In practice, only $\theta$ is observed. The variable $r$, and hence the vector $\bm{x}$ are not observable. They can be viewed as an augmented variable set. The marginal density of $\bm{\theta}$ can be obtained by integrating out $r$. The result derived by \textit{Pukkila} \& \textit{Rao} \cite{pukkila1988pattern} is given below:
\begin{align}
\begin{split}
    f(\bm{\theta}|\bm{\mu},\bm{\Sigma}) &= \int_0^\infty f(r,\bm{\theta}|\bm{\mu},\bm{\Sigma})dr\\
    &= |2\pi\bm{\Sigma}|^{-\frac{1}{2}}Q_3^{-\frac{p}{2}}\exp{\Bigl\{-\frac{1}{2}(Q_1-Q_2^2Q_3^{-1})\Bigl\}}\kappa_p(Q_2Q_3^{-\frac{1}{2}}) \label{PNpdf}
\end{split}
\end{align}
where $Q_1 = \bm{\mu}^T\bm{\Sigma}^{-1}\bm{\mu}$, $Q_2 = \bm{\mu}^T\bm{\Sigma}^{-1}\bm{u}$, $Q_3 = \bm{u}^T\bm{\Sigma}^{-1}\bm{u}$ and function $\kappa_p(\cdot)$ is defined as follows:
\begin{align*}
    \kappa_p(x) = \int_0^\infty r^{p-1}\exp{\Bigl\{-\frac{1}{2}(r-x)^2\Bigl\}}dr
\end{align*}
The recursive property of $\kappa_p(x)$ is given in \cite{pukkila1988pattern}. Equation (\ref{PNpdf}) gives the density function of $\mathcal{PN}_p(\bm{\mu},\bm{\Sigma})$. As pointed out in \cite{wang2013directional}, the shape of the distribution can be asymmetric or bimodal and the mean direction of $\bm{\theta}$ is dependent on both $\bm{\mu}$ and $\bm{\Sigma}$. If $\bm{\mu}$ is orthogonal to any of the eigenvectors of $\bm{\Sigma}$, the distribution is symmetric \cite{hernandez2017general}. 

Note that the density function remains unchanged if $\bm{\mu}$ and $\bm{\Sigma}$ are replaced by $a\bm{\mu}$ and $a^2\bm{\Sigma}$ for any $a>0$. This raises an identifiability issue that can be solved by putting a constraint on the parameters. A popular choice is to let $\bm{\Sigma} = \bm{I}_p$ which leads to a distribution that is unimodal and symmetric about the direction of $\bm{\mu}$. A more general approach is to fix one of the diagonal entries of $\bm{\Sigma}$ to be one \cite{wang2013directional,hernandez2017general,mastrantonio2018joint}:
\begin{align} \label{constraint}
\bm{\Sigma} = \begin{pmatrix} 
    1 & \bm{\omega}^T \\
    \bm{\omega} & \bm{\Omega}
    \end{pmatrix}
\end{align}
where $\bm{\omega}$ is a $(p-1)$ vector and $\bm{\Omega}$ is a $(p-1)$ by $(p-1)$ matrix. The constrained $\bm{\Sigma}$ needs to be positive semi-definite to remain a valid covariance matrix. 

\subsection{Incorporating linear variables} \label{ILV}

Suppose we observe $\bm{u}$ (or equivalently $\bm{\theta}$) together with $q$ linear variables denoted by the vector $\bm{y} = (y_1,...,y_q)^T$ that follow a multivariate normal distribution (MVN). Since $\bm{x}$, the augmented representation of $\bm{u}$, is also normally distributed, it is intuitive to introduce dependence between $\bm{u}$ and $\bm{y}$ by modeling the joint distribution of $\bm{z} = (\bm{x},\bm{y})^T$ with a $(p+q)$-variate normal:
\begin{align} \label{partitionMVN}
    \bm{z}=\begin{pmatrix} 
    \bm{x} \\
    \bm{y}
    \end{pmatrix} \sim \mathcal{N}_{d}[\Tilde{\bm{\mu}} = \begin{pmatrix}
    \bm{\mu}_x \\
    \bm{\mu}_y
    \end{pmatrix},
    \Tilde{\bm{\Sigma}} =
    \begin{pmatrix}
    \bm{\Sigma}_{xx} & \bm{\Sigma}_{xy} \\
    \bm{\Sigma}_{yx} & \bm{\Sigma}_{yy}
    \end{pmatrix}]
\end{align}
where $\Tilde{\bm{\mu}}$ and $\Tilde{\bm{\Sigma}}$ are the joint mean and covariance matrix and $d=p+q$. Note $\bm{\Sigma}_{xx}$ should satisfies the identifiability constraint (\ref{constraint}). Marginally $\bm{x}$ and $\bm{y}$ are still normally distributed. Based on the conditional distribution property of the MVN, we have $\bm{x}|\bm{y} \sim\mathcal{N}_p(\bm{\mu}_{x|y},\bm{\Sigma}_{x|y})$
where $\bm{\mu}_{x|y} = \bm{\mu}_x+\bm{\Sigma}_{xy}\bm{\Sigma}_{yy}^{-1}(\bm{y}-\bm{\mu}_y)$ and $\bm{\Sigma}_{x|y}=\bm{\Sigma}_{xx}-\bm{\Sigma}_{xy}\bm{\Sigma}_{yy}^{-1}\bm{\Sigma}_{yx}$. Substituting $(\bm{\mu}_{x|y},\bm{\Sigma}_{x|y})$ for $(\bm{\mu},\bm{\Sigma})$ in (\ref{rthetajoint}) and (\ref{PNpdf}) yields the corresponding conditional density functions for $(r,\bm{\theta})|\bm{y}$ and $\bm{\theta}|\bm{y}$. Multiplying these conditional density functions by the marginal density function of $\bm{y}$, we obtain the following joint distributions:
\begin{align}
    f(r,\bm{\theta},\bm{y}|\Tilde{\bm{\mu}},\Tilde{\bm{\Sigma}}) &= f(r,\bm{\theta}|\bm{\mu}_{x|y},\bm{\Sigma}_{x|y}) \cdot \mathcal{N}_q(\bm{y}|\bm{\mu}_{y},\bm{\Sigma}_{yy}) \label{rthetaypdf} \\
    f(\bm{\theta},\bm{y}|\Tilde{\bm{\mu}},\Tilde{\bm{\Sigma}}) &= f(\bm{\theta}|\bm{\mu}_{x|y},\bm{\Sigma}_{x|y}) \cdot \mathcal{N}_q(\bm{y}|\bm{\mu}_{y},\bm{\Sigma}_{yy}) \label{spnpdf}
\end{align}
The complete mathematical expressions for (\ref{rthetaypdf}) and (\ref{spnpdf}) are omitted here to avoid redundancy. The joint distribution of $\bm{\theta}$ and $\bm{y}$ given by (\ref{spnpdf}) is obtained by projecting a number of dimensions of a normally distributed variable onto a hypersphere, so we refer to it as the semi-projected normal distribution (SPN). It is worth noting that with $p=2$, SPN is a special case of the joint projected normal and skew-normal distribution (JPNSN) introduced in \cite{mastrantonio2018joint}. Some properties of the JPNSN still hold true for the SPN with $p>2$. For example, $\bm{\theta}$ with any subset of $\bm{y}$ is still SPN distributed; marginally $\bm{\theta}\sim\mathcal{PN}_p(\bm{\mu}_x,\bm{\Sigma}_{xx})$ and $\bm{y}\sim\mathcal{N}_q(\bm{\mu}_y,\bm{\Sigma}_{yy})$; $\bm{\Sigma}_{xy}$ describes the directional-linear dependence and $\bm{\theta}\perp\bm{y}$ if and only if $\bm{\Sigma}_{xy}=\bm{0}$.

The SPN is a very flexible distribution family, but the complex density function makes it challenging to estimate the parameters via maximum likelihood or sample from their posterior distribution. Previous studies \cite{wang2013directional,hernandez2017general,mastrantonio2018joint} exploit the close relationship between a MVN and the projected normal by augmenting the SPN with a draw of $r$ from its full conditional distribution and restoring a complete observation of $\bm{x}$ via the transformation (\ref{car2sph}). Then the posterior distribution of the MVN parameters conditional on $\bm{x}$ and $\bm{y}$ can easily be sampled from. In the case of the SPN, the full conditional of $r$ can be derived from (\ref{rthetajoint}) and (\ref{rthetaypdf}):
\begin{align}
    f(r|\bm{\theta},\bm{y},\Tilde{\bm{\mu}},\Tilde{\bm{\Sigma}}) \ \propto \  f(r,\bm{\theta},\bm{y}|\Tilde{\bm{\mu}},\Tilde{\bm{\Sigma}}) \ \propto \ 
    r^{p-1}\exp{\Bigl\{-\frac{1}{2}Q_3^*(r-\frac{Q_2^*}{Q_3^*})^2 \Bigl\}} \label{sampleR}
\end{align}
where $Q_2^* = \bm{\mu}^T_{x|y}\bm{\Sigma}_{x|y}^{-1}\bm{u}$, $Q_3^* = \bm{u}^T\bm{\Sigma}_{x|y}^{-1}\bm{u}$. We modified a slice sampling strategy proposed in \cite{hernandez2017general} to sample from (\ref{sampleR}). Details are provided in Appendix \ref{appA}.

\section{Dirichlet process mixture of semi-projected normal distributions}
\label{secdpmm}

Parametric models often encounter real-world applications for which there is insufficient prior knowledge and data to justify the parametric assumptions, or for which the parametric model is inadequate to capture the complexity of the data. Nonparametric models support a more flexible and robust specification of distributions. In the field of Bayesian nonparametrics, the Dirichlet process mixture model (DPMM) is widely used due to its elegant mathematical structure and broad applicability. In this section, we build a DPMM using the SPN as the mixture density and develop an algorithm for Bayesian inference.

\subsection{The Dirichlet process mixture model}
The basic idea of a DPMM is that an unknown density $f(\bm{z})$ can be approximated by a sum of countably infinite densities:
\begin{align} \label{dedp}
    f(\bm{z}) = \int f(\bm{z}|\bm{\phi})dG(\bm{\phi}) = \sum_{k=1}^\infty \pi_kf(\bm{z}|\bm{\phi}_k)
\end{align}
where $f(\bm{z}|\bm{\phi})$ is known as the mixture density with parameter $\bm{\phi}$, and $G$ is a discrete mixing distribution for $\bm{\phi}$ with $\pi_k$'s as probabilities. Consider a number of observations $\bm{z}_1,...,\bm{z}_n$ generated from $f(\bm{z})$. The data generation process can be expressed as follows:
\begin{align} \label{DPMM}
\begin{split}
    \bm{z}_i &\sim f(\bm{z}|\bm{\phi}_i) \\
    \bm{\phi}_i &\sim G  \\
    G &\sim DP(\alpha_0, G_0)
\end{split}
\end{align}
where $G$ is generated from a Dirichlet process \cite{ferguson1973bayesian} prior with base measure $G_0$ and concentration parameter $\alpha_0$. The Dirichlet process is a distribution on the family of distributions. With the hierarchical structure, conditional independence is implicitly assumed. For example, $\bm{z}_i$'s are independent of each other given the $\bm{\phi}_i$'s. Formulas (\ref{DPMM}) represent the most basic form of a DPMM. Additional structure can be added to the hierarchical model, e.g., putting priors on the concentration parameter $\alpha_0$ \cite{escobar1995bayesian} and base measure $G_0$ \cite{teh2006hierarchical}. 

An equivalent and more comprehensive representation of a DPMM is as the limit of a finite mixture model with the number of clusters $K$ going to infinity \cite{neal2000markov,ishwaran2002exact}:
\begin{align} \label{IMM}
\begin{split}
    \bm{z}_i | c_i,\{\bm{\varphi}_k\}_{k=1}^K  &\sim f(\bm{z}|\bm{\varphi}_{c_i}) \\
    c_i | \bm{\pi} &\sim Discrete(\pi_1,...,\pi_K) \\
    \bm{\varphi}_k &\sim G_0 \\
    \bm{\pi} &\sim Dirichlet(\alpha_0/K,...,\alpha_0/K)
\end{split}
\end{align}
In this form, $\{c_i\}_{i=1}^n$ label the cluster assignments for each observation and theoretically can take any $K$ distinct values (integers 1 to $K$ are used here) and the relevant parameters for observation $\bm{z}_i$ is $\bm{\phi}_i = \bm{\varphi}_{c_i}$. The probabilities $\bm{\pi}=(\pi_1,...,\pi_K)$ indicate how likely it is that a new observation will be assigned to each of the clusters. The two representations of the DPMM given in (\ref{DPMM}) and (\ref{IMM}) correspond to its two most popular applications: density estimation and clustering.

Bayesian inference for the DPMM mainly involves sampling from the posterior distribution of $\{\bm{\phi}_i\}_{i=1}^n$ and $\{c_i\}_{i=1}^n$ by simulating a Markov chain that reaches equilibrium at that distribution. \textit{Neal} \cite{neal2000markov} provides several Gibbs sampling algorithms for DPMMs. When $G_0$ is a conjugate prior distribution for $f(\bm{z}|\bm{\phi})$, the collapsed Gibbs sampler (\textbf{Algorithm 3} in \cite{neal2000markov}) has a better convergence rate than other sampling algorithms \cite{maceachern1994estimating}. The algorithm directly samples $\{c_i\}_{i=1}^n$ without updating $\{\bm{\phi}_i\}_{i=1}^n$. Each Gibbs sampling iteration consists of assigning each $\bm{z_i}$ to an existing cluster or a new one by evaluating the full conditional distribution of $c_i$ given all $c_j$ but $c_i$ (written as $c_{-i}$):
\begin{align} \label{cpost}
    P(c_i=k|c_{-i}, \bm{z}_i)\ \propto
    \begin{cases}
    n_{-i,k}\int f(\bm{z}_i|\bm{\phi})dG_{-i,k}(\bm{\phi})\qquad &\text{if }k\text{ represents an existing cluster}  \\
    \alpha_0\int f(\bm{z}_i|\bm{\phi})dG_0(\bm{\phi}) \qquad &\text{if }k\text{ represents a new cluster}
    \end{cases}
\end{align}
Here $n_{-i,k}$ is the number of $c_j$ for $j\neq i$ that are equal to $k$, and $G_{-i,k}$ is the posterior distribution of $\bm{\phi}$ based on $G_0$ and all observations $\bm{z}_j$ for which $j\neq i$ and $c_j = k$. Evaluation of the integrals in (\ref{cpost}) becomes much simpler when $G_0$ is a conjugate prior distribution for $f(\bm{z}|\bm{\phi})$. In that case, $G_{-i,k}$ will be in the same distribution family as $G_0$, and therefore all integrals can be viewed as marginal distributions of $\bm{z}_i$ given different prior parameters.

\subsection{Incorporating the semi-projected normal distribution}

Directly using the SPN as the mixture distribution $f(\bm{z}|\bm{\phi})$ in a DPMM can be challenging due to the complexity of its density function and the lack of a conjugate prior distribution. Instead, we choose to model the complete (augmented) data $\bm{z}=(\bm{x},\bm{y})^T$ with a MVN likelihood as shown in (\ref{partitionMVN}). In this case, since $\bm{\phi} = (\Tilde{\bm{\mu}}, \Tilde{\bm{\Sigma}})$, it is natural to use the normal inverse-Wishart distribution as a conjugate prior. However, as mentioned in Section \ref{SPN}, the covariance matrix $\Tilde{\bm{\Sigma}}$ needs to satisfy the identifiability constraint that at least one of the diagonal entries equals one. The inverse-Wishart distribution does not satisfy that constraint and hence can not be directly applied. \textit{Hernandez-Stumpfhauser et al.} \cite{hernandez2017general} provided a reparametrization for the covariance matrix of the projected normal distribution to ensure its positive semi-definiteness and allow separate prior distributions on the constituent submatrices.

We propose using a conditional inverse-Wishart distribution to accommodate the constraint. Suppose $\Tilde{\bm{\Sigma}}$ follows an inverse-Wishart distribution $\mathcal{IW}(\bm{S},\nu)$. Partition $\Tilde{\bm{\Sigma}}$ and $\bm{S}$ conformably with each other:
\begin{align} \label{partitionS}
    \Tilde{\bm{\Sigma}} = \begin{pmatrix}
    \bm{\Sigma}_{11} & \bm{\Sigma}_{12} \\
    \bm{\Sigma}_{21} & \bm{\Sigma}_{22}
    \end{pmatrix},\quad
    \bm{S} = \begin{pmatrix}
    \bm{S}_{11} & \bm{S}_{12} \\
    \bm{S}_{21} & \bm{S}_{22}
    \end{pmatrix}
\end{align}
Here $\bm{\Sigma}_{ij}$ and $\bm{S}_{ij}$ are $d_i\times d_j$ matrices (with $d_1+d_2=d=p+q$ and $d_1\leq p$) and satisfy the following properties:
\begin{align} \label{CIWprop}
\begin{split}
    &\text{(a) }\bm{\Sigma}_{11}\sim\mathcal{IW}(\bm{S}_{11}, \nu - d_2)  \\
    &\text{(b) }\bm{\Sigma}_{11} \text{ is independent of } \bm{\Sigma}_{11}^{-1}\bm{\Sigma}_{12} \text{ and } \bm{\Sigma}_{22\cdot1} \text{, where } \bm{\Sigma}_{22\cdot1}=\bm{\Sigma}_{22}-\bm{\Sigma}_{21}\bm{\Sigma}_{11}^{-1}\bm{\Sigma}_{12} \\
    &\text{(c) }\bm{vec}(\bm{\Sigma}_{11}^{-1}\bm{\Sigma}_{12})| \bm{\Sigma}_{22\cdot1} \sim \mathcal{N}_{d_1\times d_2}[\bm{vec}(\bm{S}_{11}^{-1}\bm{S}_{12}), \bm{\Sigma}_{22\cdot1} \otimes \bm{S}_{11}^{-1}] \\
    &\text{(d) }\bm{\Sigma}_{22\cdot1} \sim \mathcal{IW}(\bm{S}_{22\cdot1}, \nu) \text{, where } \bm{S}_{22\cdot1}=\bm{S}_{22}-\bm{S}_{21}\bm{S}_{11}^{-1}\bm{S}_{12}
\end{split}
\end{align}
The operator $\bm{vec}(\cdot)$ vectorizes a matrix by stacking its columns on top of one another, and $\otimes$ is the Kronecker product. In our case, $d_1$ is at least one to ensure identifiability but can be larger (e.g., $d_1=p$). The proof of (\ref{CIWprop}) is provided in Appendix \ref{appB}. The properties listed above suggest a reparameterization of $\Tilde{\bm{\Sigma}}$ as $(\bm{\Sigma}_{11},\bm{\Sigma}_{11}^{-1}\bm{\Sigma}_{12},\bm{\Sigma}_{22\cdot 1})$. We can derive the conditional distribution of $(\bm{\Sigma}_{11}^{-1}\bm{\Sigma}_{12},\bm{\Sigma}_{22\cdot 1}|\bm{\Sigma}_{11})$ using properties (b) - (d) as:
\begin{align} \label{CIWpdf}
\begin{split}
    f(\bm{\Sigma}_{11}^{-1}\bm{\Sigma}_{12},\bm{\Sigma}_{22\cdot 1}|\bm{\Sigma}_{11}) &= f(\bm{\Sigma}_{11}^{-1}\bm{\Sigma}_{12},\bm{\Sigma}_{22\cdot 1}) \\
    &= f(\bm{\Sigma}_{11}^{-1}\bm{\Sigma}_{12}|\bm{\Sigma}_{22\cdot 1})f(\bm{\Sigma}_{22\cdot 1}) \\
    &= \mathcal{N}_{d_1\times d_2}[\bm{vec}(\bm{S}_{11}^{-1}\bm{S}_{12}), \bm{\Sigma}_{22\cdot1} \otimes \bm{S}_{11}^{-1}] \cdot \mathcal{IW}(\bm{S}_{22\cdot1}, \nu)
\end{split}
\end{align}
If $\bm{\Sigma}_{11}$ is fixed as constant, equation (\ref{CIWpdf}) provides a distribution to sample the rest of $\Tilde{\bm{\Sigma}}$ and allows us to evaluate the likelihood of the sample. And $\Tilde{\bm{\Sigma}}$ sampled from (\ref{CIWpdf}) is bound to be positive definite if $\bm{\Sigma}_{11}$ is positive definite. We call this distribution the conditional inverse-Wishart (CIW). The CIW perfectly fits our demand to constrain a part of the covariance matrix ($\bm{\Sigma}_{11}$). We can either choose $\bm{\Sigma}_{11}$ equal one ($d_1 = 1$) to satisfy the constraint (\ref{constraint}) and provide a flexible distribution; or let $\bm{\Sigma}_{11} = \bm{\Sigma}_{xx} = \bm{I}_p$ ($d_1 = p$) such that the marginal of $\bm{\theta}$ is unimodal and symmetrical about $\bm{\mu}_x$.

The inverse-Wishart distribution is often used as a conjugate prior for the covariance matrix. Based on the reparameterization of $\Tilde{\bm{\Sigma}}$ above, the inverse-Wishart can be expressed as the product of a CIW distribution conditional on $\bm{\Sigma}_{11}$ and the marginal of $\bm{\Sigma}_{11}$. Therefore, the CIW is also a conjugate prior to $\Tilde{\bm{\Sigma}}$ when $\bm{\Sigma}_{11}$ is fixed. Assume observations $\bm{z}_1,...,\bm{z}_n\overset {iid} \sim \mathcal{N}_d(\Tilde{\bm{\mu}},\Tilde{\bm{\Sigma}})$ and the following priors on $(\Tilde{\bm{\mu}},\Tilde{\bm{\Sigma}})$:
\begin{align}
\begin{split}
    \Tilde{\bm{\mu}} | \Tilde{\bm{\Sigma}} &\sim \mathcal{N}_d(\bm{\mu}_0,\frac{1}{\lambda_0}\Tilde{\bm{\Sigma}}) \\
    \Tilde{\bm{\Sigma}} &\sim \mathcal{CIW}(\bm{S}_0,\nu_0) \\
\end{split}
\end{align}
Let $\mathcal{NCIW}(\bm{\Psi}_0)$ denote the joint distribution formed by these priors where $\bm{\Psi}_0 = (\bm{\mu}_0,\lambda_0,\bm{S}_0,\nu_0)$ is the set of hyperparameters. Then the posterior distribution follows $\mathcal{NCIW}(\bm{\Psi}_n)$ with $\bm{\Psi}_n = (\bm{\mu}_n,\lambda_n,\bm{S}_n,\nu_n)$  defined as:
\begin{equation} \label{postpara}
\begin{gathered}
\bm{\mu}_n = \frac{\lambda_0\bm{\mu}_0+n\Bar{\bm{z}}}{\lambda_0+n} \\
\lambda_n = \lambda_0+n \\
\nu_n = \nu_0+n \\
\bm{S}_n = \bm{S}_0 + \sum_{i=1}^n\bm{z}_i\bm{z}_i^T - (\lambda_0+n)\bm{\mu}_n\bm{\mu}_n^T+\lambda_0\bm{\mu}_0\bm{\mu}_0^T
\end{gathered}
\end{equation}
where $\Bar{\bm{z}}$ is the sample mean. The conjugacy of the prior allow us to directly sample the cluster assignments $\{c_i\}_{i=1}^{n}$ from (\ref{cpost}). Here $G_0$ is $\mathcal{NCIW}(\bm{\Psi}_0)$ and $G_{-i,c}$ also follows the $\mathcal{NCIW}$ with parameters derived according to (\ref{postpara}) for cluster $c$. If we fix $\bm{\Sigma}_{11} = \bm{I}_{d_1}$, the marginal distribution for the data $\bm{z}_1,...,\bm{z}_n$ can be derived based on the conjugacy and Bayes rule:
\begin{align} \label{datamarginal}
\begin{split}
    f(\bm{z}_1,...,\bm{z}_n|\bm{\Psi}_0) 
    &= \frac{f(\bm{z}_1,...,\bm{z}_n|\Tilde{\bm{\mu}},\Tilde{\bm{\Sigma}})\times f(\Tilde{\bm{\mu}},\Tilde{\bm{\Sigma}}|\bm{\Psi}_0)}{f(\Tilde{\bm{\mu}},\Tilde{\bm{\Sigma}}|\bm{z}_1,...,\bm{z}_n,\bm{\Psi}_0)} \Bigl|_{\Tilde{\bm{\mu}}=\bm{0},\Tilde{\bm{\Sigma}}=\bm{I}_d} \\
    &= \frac{\prod_{i=1}^n \mathcal{N}_d(\bm{z}_i|\bm{0},\bm{I}_d)\cdot\mathcal{NCIW}(\bm{0},\bm{I}_d|\bm{\Psi}_0)}{\mathcal{NCIW}(\bm{0},\bm{I}_d|\bm{\Psi}_n)} \\
    &=
    \Bigl[2^{nd_1}\pi^{nd}(\frac{\lambda_n}{\lambda_0})^d\frac{|\bm{S}_n|^{\nu_n}}{|\bm{S}_0|^{\nu_0}}\frac{{|\bm{S}_n}_{11}|^{d_2-\nu_n}}{|{\bm{S}_0}_{11}|^{d_2-\nu_0}}\exp{\Bigl\{\bm{tr}({\bm{S}_n}_{11}-{\bm{S}_0}_{11})\Bigl\}}\Bigl]^{-\frac{1}{2}}\prod_{j=1}^{d_2}\frac{\Gamma(\frac{\nu_n+1-j}{2})}{\Gamma(\frac{\nu_0+1-j}{2})} 
\end{split}
\end{align}
where ${\bm{S}_0}_{11}$ and ${\bm{S}_n}_{11}$ are submatrices of $\bm{S}_0$ and $\bm{S}_n$ partitioned according to (\ref{partitionS}), and $\Gamma(\cdot)$ is the gamma function. The integrals in (\ref{cpost}) are special cases of (\ref{datamarginal}) and hence can be directly calculated.

For the rest of the paper, we refer to our method using the acronym DPSPN to indicate the Dirichlet process semi-projected normal mixture model. Algorithm \ref{algorithm} provides the pseudocode to sample from the DPSPN using a Gibbs sampler. The initialization of $\{c_i\}_{i=1}^n$ and $\{r_i\}_{i=1}^n$ can incorporate prior knowledge and preprocessing results from other algorithms. In this study, we initialize the algorithm by randomly grouping the data into different clusters and sampling the radius of each observation from an exponential distribution with parameter 1.
\begin{algorithm} 
\caption{Gibbs Sampler for the DPSPN}\label{algorithm}
\begin{algorithmic}
\State Random initialization of $\{c_i\}_{i=1}^n$, and $\{r_i\}_{i=1}^n$
\State $K = \#$ of clusters 
\For{$iter = 1$ to $M$}
    \State update $\{x_i\}_{i=1}^n$ with $\{r_i\}_{i=1}^n$ using (\ref{car2sph})
    \For{$i = 1$ to $n$}
        \State remove $z_i$ from its current cluster $c_i$
        \State update the posterior parameter $\bm{\Psi}$ of cluster $c_i$ using (\ref{postpara})
        \State if the cluster is empty, remove it and decrease $K$
        \For{$k = 1$ to $K$}
            \State calculate $P(c_i=k|c_{-i},\bm{z}_i)\ \propto\  n_{-i,k}f(\bm{z}_i|\bm{\Psi}^k)$ using (\ref{datamarginal}) \Comment{$\bm{\Psi}^k$ is the hyperparameter for cluster $k$}
        \EndFor
        \State calculate $P(c_i=k^*|c_{-i},\bm{z}_i)\ \propto\  \alpha_0 f(\bm{z}_i|\bm{\Psi}_0)$ using (\ref{datamarginal}) \Comment{$k^*$ is a new cluster}
        \State sample a new value for $c_i$ from $P(c_i|c_{-i},\bm{z}_i)$ after normalizing the above probabilities
        \State add $\bm{z_i}$ to cluster $c_i$
        \State update $\bm{\Psi}^{c_i}$ using (\ref{postpara})
        \State if a new cluster is created (i.e., $c_i=k^*$ was selected), increase $K$
    \EndFor
    \For{$k = 1$ to $K$}
        \State sample $(\Tilde{\bm{\mu}}^k,\Tilde{\bm{\Sigma}}^k)$ from $\mathcal{NCIW}(\bm{\Psi}^k)$
    \EndFor
    \For{$i = 1$ to $n$}
        \State sample $r_i$ from $f(r|\bm{\theta}_i,\bm{y}_i,\Tilde{\bm{\mu}}^{c_i},\Tilde{\bm{\Sigma}}^{c_i})$ using (\ref{sampleR})
    \EndFor
    \State update the concentration parameter $\alpha_0$ (optional, see \cite{escobar1995bayesian})
\EndFor
\end{algorithmic}
\end{algorithm}

\section{Clustering Experiments}
\label{Experiments}

We implemented the DPSPN in C++ by modifying a DPMM package \cite{Märtens2018} and posted the source code on GitHub (\url{https://github.com/zout3/DPSPN}). In this section, Our model is tested in an experiment clustering synthetic data and in a real world application to image segmentation and compared with methods introduced in other studies. In all of the situations, we use a non-informative proper hyperprior distribution by setting the hyperparameters as follows:
\begin{gather}\label{priorpara}
    \bm{\mu}_0 = \bm{0}, \ \lambda_0 = 1, \ \nu_0 = d + 2,\ \bm{S}_0 = \bm{I}_d, \text{ and } \alpha_0 = 1
\end{gather}

\subsection{Synthetic data}

\begin{figure}
  \centering
  \resizebox{1\textwidth}{!}{%
    \includegraphics{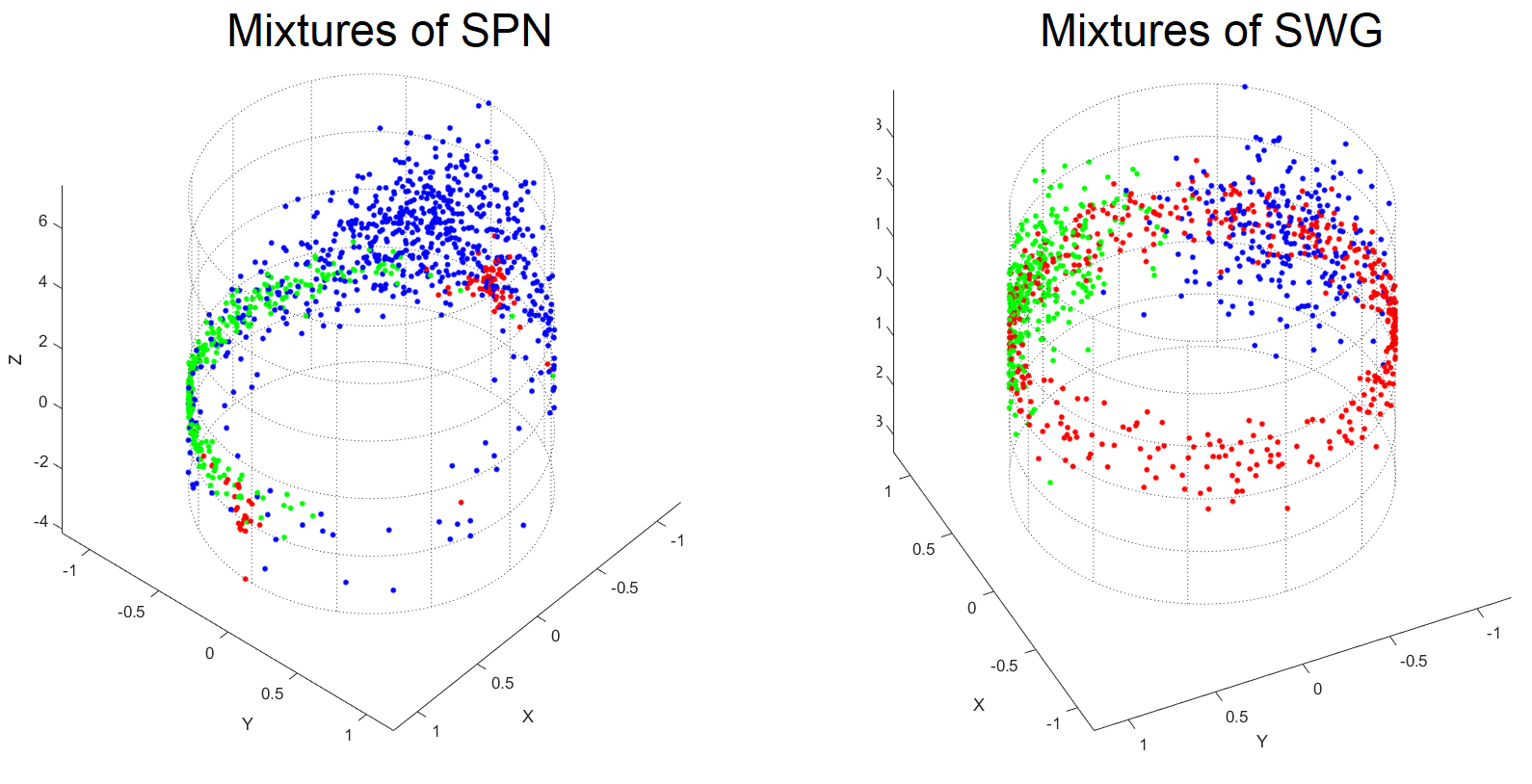}
    }
  \caption{Examples of mixture data of SPN (left) and SWG (right). Both plots contain 1000 data points with three clusters denoted in colors of blue, red and green.}
  \label{SWGSPN}
\end{figure}

The synthetic data are generated from the finite mixture model defined in (\ref{IMM}). We use the MVN as the mixture density $f(\bm{z}|\bm{\phi})$ and the normal inverse-Wishart distribution as the base measure $G_0$. The hyperparameters of $G_0$ are given in (\ref{priorpara}). The directional-linear data can be obtained by either projecting the first $p$ dimensions into $\mathbb{S}^{p-1}$, or taking the modulo of the first dimension over $2\pi$. The first case is exactly the SPN distribution. The second case only yields circular-linear data and is called the semi-wrapped Gaussian (SWG) in \cite{roy2016swgmm}. With the same sample space, the SWG consists of fewer parameters than the SPN and thus has less flexibility. For example, the marginal of the SWG is the wrapped normal distribution which is always unimodal and symmetric. Both approaches are applied here to simulate data with one circular dimension ($p=2$ for SPN and $p=1$ for SWG) and one linear dimension ($q=1$). Sample datasets are displayed in Figure \ref{SWGSPN}. The flexibility of the SPN can be observed here in the asymmetric shape of the blue cluster and the bimodal shape of the red cluster.

To gradually increase the data complexity, the number of clusters $K$ is varied from 2 to 8. For each $K$, datasets composed of 1000 data points are generated following the procedure described above. Since the cluster parameters are highly variable due to the non-informative prior, we analyze 100 datasets for each choice of $K$. We then fit the model to the simulated datasets independently and acquire an average estimate of model performance under level $K$. In terms of the covariance matrix constraint, the DPSPN is applied with both $\bm{\Sigma}_{11} = 1$ and $\bm{\Sigma}_{11} = I_2$ to demonstrate the different degrees of flexibility provided.

For each simulated dataset, we run the Gibbs sampler to produce 4 Monte Carlo chains with different initializations. Each chain is iterated 6000 times and the first 5000 samples are eliminated as a burn-in period to obtain 1000 draws from the posterior distribution. For convergence diagnosis, we compute the Gelman–Rubin statistic \cite{gelman1992inference,brooks1998general} based on the likelihood of the complete data given in (\ref{datamarginal}) over the 4000 posterior clustering (1000 clustering from each of the four chains), and obtain values of the statistic smaller than 1.2 for all datasets.

To simplify the evaluation, we adopt the \text{SALSO} algorithm proposed by \textit{Dahl et al.} \cite{dahl2022search} to produce a consensus clustering as a summary of the posterior distribution of data clusterings. Assume $\bm{C}_1,...,\bm{C}_{N}$ ($N=4000$) are the posterior clusterings obtained from Gibbs sampling for a given dataset. The consensus clustering $\bm{C}^*$ can be estimated as:
\begin{gather} \label{salso}
    \bm{C}^* = \text{argmin}_C\sum_{i=1}^N\text{VoI}(\bm{C},\bm{C}_i)
\end{gather}
where $\text{VoI}(\cdot,\cdot)$ denotes the variation of information \cite{meilua2007comparing}
that measures the distance between two clusterings. More formally, the VoI of two clusterings $\bm{C}_1$ (with $K_1$ clusters) and $\bm{C}_2$ (with $K_2$ clusters) of $n$ observations is defined as the sum of their entropies $\text{H}(\bm{C}_1)$ and $\text{H}(\bm{C}_2)$ minus twice their mutual information $\text{MI}(\bm{C}_1,\bm{C}_2)$:
\begin{align} \label{voidef}
    \text{VoI}(\bm{C}_1,\bm{C}_2) &= \text{H}(\bm{C}_1) + \text{H}(\bm{C}_2) - 2\text{MI}(\bm{C}_1,\bm{C}_2) \\
    &= \sum_{i=1}^{K_1}\frac{n_i}{n}\log_2(\frac{n}{n_i}) + \sum_{j=1}^{K_2}\frac{n_j}{n}\log_2(\frac{n}{n_j}) - 2\sum_{i=1}^{K_1}\sum_{j=1}^{K_2}\frac{n_{ij}}{n}\log_2(\frac{n_{ij}n}{n_in_j})
\end{align}
Here $n_i$ and $n_j$ are respectively the numbers of observations in cluster $i$ of $\bm{C}_1$ and in cluster $j$ of $\bm{C}_2$, and $n_{ij}$ is the number of observations in both cluster $i$ of $\bm{C}_1$ and cluster $j$ of $\bm{C}_2$.

To evaluate the model performance on data clustering, the adjusted Rand index (ARI) \cite{hubert1985comparing} is applied to measure the discrepancy between the consensus clustering $\bm{C}^*$ given by the model and the ground truth (known for simulated data). The Rand index \cite{rand1971objective} is a measure of the similarity between two clusterings. Given a
set of $n$ elements, the Rand index between two clusterings $\bm{C}_1$ and $\bm{C}_2$ is computed as follows:
\begin{gather} \label{RIdef}
    RI(\bm{C}_1,\bm{C}_2) = \frac{a+b}{\begin{pmatrix}
    n \\
    2
    \end{pmatrix}}
\end{gather}
where $a$ is the number of pairs of elements that are placed in the same cluster in $\bm{C}_1$ and in the same cluster in $\bm{C}_2$ , and $b$ is the number of pairs placed in different clusters in $\bm{C}_1$ and in different clusters in $\bm{C}_2$. The ARI is a modified version that corrects the clustering similarity measure for chance agreement under the permutation model \cite{xuan2010information}:
\begin{gather}
    ARI(\bm{C}_1,\bm{C}_2) = \frac{RI(\bm{C}_1,\bm{C}_2)-\mathbb{E}[RI(\bm{C}_1,\bm{C}_2)]}{1-\mathbb{E}[RI(\bm{C}_1,\bm{C}_2)]}
\end{gather}

\begin{figure}
  \centering
  \resizebox{1\textwidth}{!}{%
    \includegraphics{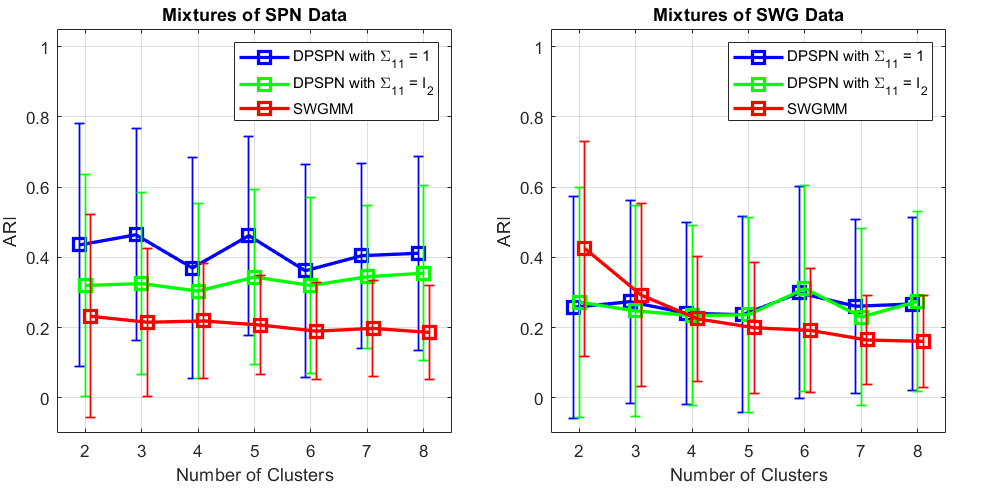}
    }
  \caption{Clustering results from DPSPN and SWGMM on mixture data of SPN (left) and SWG (right) with different numbers of clusters. Each point is an average of ARI over 100 datasets and the error bar denotes one standard deviation above and below the average.}
  \label{ARIplot}
\end{figure}

We compare our model with the SWGMM proposed in \cite{roy2016swgmm}. The SWGMM prespecifies the number of clusters $K$ and uses SWG as the mixture distribution and therefore can fit circular-linear data. To apply the SWGMM, the data is first preprocessed by a $K$-means clustering algorithm as an initialization. Then an EM algorithm is iterated 100 times to update the model parameters. For our study, we apply the SWGMM using the true number of simulated clusters $K$. Figure \ref{ARIplot} shows the clustering results on the synthetic data. The overall value of ARI is not very high due to the random data generating process where clusters are not always well-separated, hence more challenging for the model. For data generated from SPN, the DPSPN is consistently better than the SWGMM, and the more flexible version of DPSPN ($\bm{\Sigma}_{11} = 1$) is consistently better than the simpler version ($\bm{\Sigma}_{11} = \bm{I}_2$). For data generated from SWG, the SWGMM performs better when the number of clusters is low. The difference becomes less significant among the three models as the number of clusters increases, and the DPSPN has a higher ARI average for five or more clusters. The results show that DPSPN is a more flexible model than SWGMM, but it can also fit well to simpler data forms like those generated by the SWG.

\subsection{Image segmentation}

Image segmentation has become increasingly important due to its applications in many computer vision tasks like object detection and recognition \cite{wang2007object} and in medical imaging \cite{pham2000survey,roy2017jclmm}. Image segmentation can be viewed as a clustering problem that involves partitioning the image into different groups of objects according to the color of each pixel. Besides the RGB (red, green, blue) color space, many other color spaces can be used to represent images. The LUV is a special color space in which Euclidean distance provides a perceptually uniform spacing of colors \cite{kato2006markov}. Due to this property some studies have adopted the LUV space for image segmentation \cite{shafarenko1997automatic,kato2006markov,mignotte2008segmentation}. A cylindrical representation of the LUV space is to transform the U,V plane to polar coordinates and address the radial distance as chroma C and the angle as hue H. With the lightness L unchanged, this representation is known as the HCL (or LCH) space \cite{ihaka2003colour}. Since lightness and chroma are linear variables, and hue is a circular variable, image segmentation in the HCL space is equivalent to clustering directional-linear data. Compared to using linear models (e.g., a Gaussian mixture model) to cluster in the LUV space, one advantage of using the DPSPN in the HCL space is that the shape of a cluster can be more variable than in the completely linear case due to there being more model parameters.

For our experiment, we consider the Berkeley image database (BSD300) \cite{martin2001database} that has been widely used to benchmark image segmentation algorithms. The data set contains 300 color images with size 481x321. Each image was presented to multiple human subjects to perform manual segmentation. These manual segmentations are used as ground truth to evaluate the performance of a segmentation algorithm. A number of metrics are frequently adopted to assess the quality of an image segmentation. The probabilistic Rand index (PRI) \cite{unnikrishnan2005measure} is similar to the Rand index defined in (\ref{RIdef}) but instead estimates the probability that an arbitrary pair of pixels has consistent labels in two clusterings. The variation of information (VoI) \cite{meilua2007comparing} as given in (\ref{salso}) defines a distance metric between two clusterings defined by mutual information and entropy. VoI measures the amount of randomness in one segmentation which cannot be explained by the other. The global consistency error (GCE) \cite{martin2001database} measures the degree to which one clustering can be considered as a refinement of the other. An issue with GCE is that it does not penalize oversegmentation (each pixel having its own cluster achieves zero error). The boundary displacement error (BDE) \cite{freixenet2002yet} measures the average displacement error of boundary pixels between two segmented images. More precisely, the error of one boundary pixel is defined as the distance between the pixel and the closest pixel in the
other boundary image. Except for PRI, metrics with smaller values indicate better performance. As noted in \cite{yang2008unsupervised}, evaluating clustering performance using the PRI and VoI seems to better correspond with human visual perception. \textit{Roy et al.} \cite{roy2016swgmm} applied several clustering algorithms to the BSD300 and reported benchmark performance based on the metrics discussed above. The algorithms compared include several mixture models with circular-linear distributions and hence make appropriate comparators for the DPSPN.

\begin{figure}
  \centering
  \resizebox{1\textwidth}{!}{%
    \includegraphics{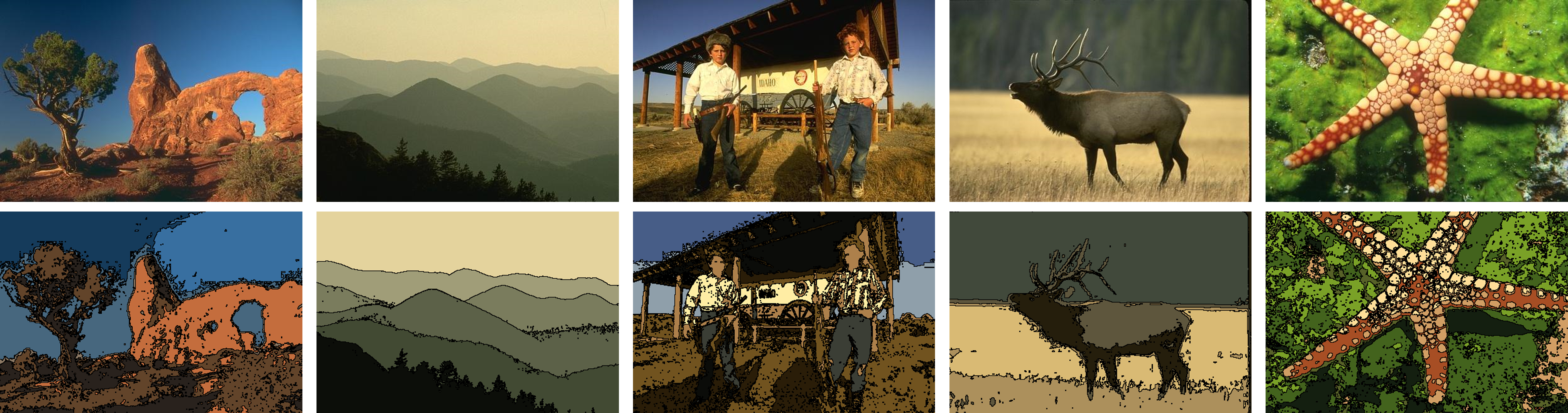}
    }
  \caption{Examples of image segmentation by the DPSPN. Original images are shown in the top row. The corresponding segmented images are shown in the bottom row.}
  \label{imgseg}
\end{figure}

We convert the images of the BSD300 into LCH color space and apply the DPSPN to each image. For each image, we run 4 chains of Gibbs sampling with each sampler iterated 6000 times. The first 5000 iterations are eliminated as a burn-in period and the remaining 1000 are thinned by keeping every 4th iteration. The Gelman-Rubin statistic computed are below 1.2 for each image. Given the resulting 1000 posterior clusterings, the SALSO algorithm \cite{dahl2022search} produces a posterior consensus clustering of the image. Figure \ref{imgseg} gives a few examples of images and their DPSPN segmentations. Oversegmentation can be observed in the background of some images. This is due to the fact that our approach does not explicitly use any spatial information in order to compare the segmentation performance with that of other circular-linear models provided in \cite{roy2016swgmm}. Another potential cause is the label switching problem \cite{stephens2000dealing} that happens in Bayesian inference. The SALSO algorithm can partially reduce the label switching effect by averaging over multiple posterior clusterings.

The four metrics are computed to quantitatively evaluate the results. Notice that every image has a number of different human segmentations as potential ground truth. The metrics are averaged across the multiple comparisons for each image. Table \ref{imgsegrslt} shows the mean value of the metrics over 300 images obtained from the DPSPN along with results of the other models reported in \cite{roy2016swgmm}. The GMM and BMM \cite{roy2007beta} are mixture models of MVN and multivariate beta distributions that are applied to the LUV space. The IvMGMM and IvMBMM \cite{roy2012mixture} are mixture models of von Mises Gaussian and von Mises Beta distributions that are applied to the LCH space. The DMM \cite{boutemedjet2008hybrid} is the mixture model of generalized Dirichlet distributions applied to the RGB space. The DPSPN outperforms the other models in terms of PRI, VoI and BDE. It has the second lowest GCE score. There results demonstrate the flexibility and excellent clustering provided by the DPSPN.

\begin{table}[ht]
\centering
\begin{threeparttable}
 \caption{Evaluation metrics of image segmentation on the BSD300 for different models}
  \label{imgsegrslt}
  \begin{tabular}{ccccc}
 \toprule
 Models & PRI & VoI & GCE & BDE \\ 
 \toprule
 DPSPN & \textbf{0.7287} & \textbf{2.5881} & 0.3324 & \textbf{14.6192} \\ 
 SWGMM$^*$ & 0.7223 & 2.6998 & 0.3486 & 15.2806 \\
 GMM$^*$ & 0.7040 & 2.8786 & 0.3608 & 15.9192 \\
 BMM$^*$ & 0.7014 & 2.8725 & 0.3688 & 15.8855 \\
 DMM$^*$ & 0.6302 & 2.8232 & \textbf{0.3241} & 17.0081 \\
 IvMGMM$^*$ & 0.7058 & 2.9117 & 0.3773 & 15.9616 \\
 IvMBMM$^*$ & 0.6494 & 2.9763 & 0.3616 & 20.4416 \\
 \toprule
\end{tabular}
\begin{tablenotes}
      \small
      \item $^*$ Results are obtained from \cite{roy2016swgmm}.
    \end{tablenotes}
\end{threeparttable}
\end{table}

\section{Bloodstain pattern analysis}
\label{BPA}

The development of the DPSPN was motivated by the desire to provide improved methods for the analysis for bloodstain pattern evidence found at crime scenes. A bloodstain pattern is a collection of stains observed at a crime scene. The main objective for bloodstain pattern analysis (BPA) is to determine the causal mechanism behind the bloodletting event \cite{damelio2001bloodstain}. By analyzing the shapes, sizes, orientations and locations of bloodstains along with other information, BPA experts develop hypotheses about how the event may have happened. Recent studies \cite{national2009strengthening,hicklin2021accuracy} have noted the subjectivity of the approach and spurred research on alternative approaches. Some research works have been done on the development of quantitative method to assess hypotheses regarding the cause of bloodstain patterns \cite{arthur2018automated,liu2020automatic,zou2022towards}. In these studies, the bloodstains are first approximated by ellipses, and then features are designed based on the parameters of the ellipses for further analysis. \textit{Arthur et al.} \cite{arthur2018automated} and \textit{Liu et al.} \cite{liu2020automatic} frame the question as a classification problem between two specified mechanisms. \textit{Zou et al.} \cite{zou2022towards} proposed the use of the likelihood ratio (LR) to measure the strength of the evidence supporting one hypothesis against another. Given a bloodstain pattern $\bm{p}$, let $H_1$ and $H_2$ denote two competing hypothesis regarding the bloodletting mechanism. The LR of evidence $\bm{p}$ regarding the two hypotheses can be written as:
\begin{gather} \label{LRdef}
    LR = \frac{f(\bm{p}|H_1)}{f(\bm{p}|H_2)}
\end{gather}
where $f(\bm{p}|H)$ is the likelihood of pattern $\bm{p}$ assuming $H$ is the true causal mechanism. The LR approach can be generalized to consider multiple hypotheses. In \cite{zou2022towards} the likelihood of a bloodstain pattern is approximated by the likelihood of a small number of features. However, a limitation of all feature-based approaches is the inevitable loss of information. The distribution of the bloodstains (ellipses) in the pattern is summarized by some features that may not be useful in distinguishing between different hypotheses. In addition, the features are case-dependent and often need redesigning for a different scenario.

We consider a different approach to estimate the likelihood of a bloodstain pattern. A bloodstain approximated by an ellipse can be represented by its five parameters $\bm{e} =(\theta,y_1,y_2,y_3,y_4)$, where $\theta$ is the angle between the x-axis and the major axis of the ellipse, and the linear component $\bm{y}=(y_1,y_2,y_3,y_4)$ are the the center coordinates $(y_1,y_2)$ of the ellipse relative to the center of the pattern and the radii of major and minor axes $(y_3,y_4)$ of the ellipse. Then we can view a bloodstain pattern $\bm{p} = (\bm{e}_1,...,\bm{e}_n)$ as a collection of quintuples. Assuming these quintuples are independent and identically distributed from a five dimensional density $f_{\bm{p}}(\bm{e})$, then the likelihood of $\bm{p}$ is $\prod_{i=1}^n f_{\bm{p}}(\bm{e}_i)$. Obtaining the likelihood of a pattern requires estimating $f_{\bm{p}}(\bm{e})$. Since the slope of an ellipse $\theta$ is a circular variable and $\bm{y}$ are all linear variables, the DPSPN can be use in this application. Here we apply the density estimation perspective of the DPMM as shown in the model specification (\ref{dedp}). 

\begin{figure}[ht]
  \centering
  \resizebox{0.6\textwidth}{!}{%
    \includegraphics{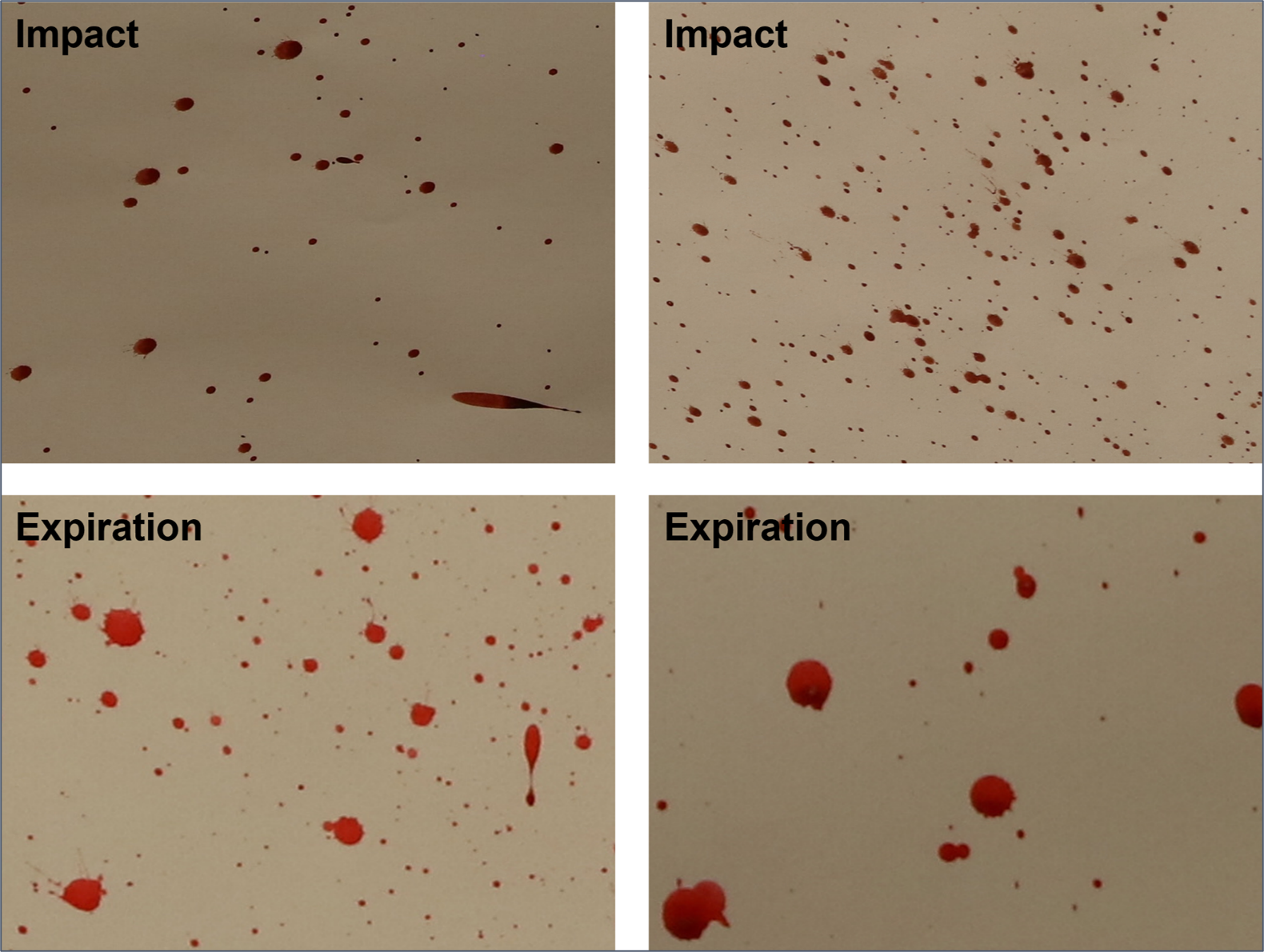}
    }
  \caption{Examples of impact patterns (the first row) and expiration patterns (the second row).}
  \label{bloodpattern}
\end{figure}

Two sets of bloodstain pattern images provided by the \textit{Institute of Environmental Science and Research, New Zealand} are used for this experiment. All patterns were generated in the laboratory with swine blood and collected on a vertical cardboard sheet. One set contains 172 impact patterns that were created by releasing a metal cylinder at some height above a blood pool, which simulates stepping into a puddle of blood. The other set contains 112 expiration patterns created by researchers coughing, speaking, shouting and spitting blood onto the target board. All patterns are scanned into image format at a resolution of 300dpi. Figure \ref{bloodpattern} shows some examples of the bloodstain patterns. We applied the technique from the work of \textit{Zou et al.} \cite{zou2021recognition} to represent each pattern $\bm{p}_j$ with a collection of ellipses $(\bm{e}_{j1},...,\bm{e}_{jn_j})$. It can be a challenging task to differentiate impact patterns from expiration patterns for BPA experts as shown by examples in the recent black box study \cite{hicklin2021accuracy}. Based on the available data, we set $H_1$ and $H_2$ regarding a bloodstain pattern as the following:
\begin{gather*}
    H_1:\text{The pattern is caused by impact.} \qquad vs \qquad
    H_2:\text{The pattern is caused by expiration.} 
\end{gather*}
Our strategy is to build a data-driven model that can train on a set of patterns with known causal mechanism. The DPSPN can only estimate the density function $f_{\bm{p}_j}(\bm{e})$ of a single bloodstain pattern $\bm{p}_j$ at one time. To address variation in patterns from the same mechanism, we need to extend the DPMM to a hierarchical Dirichlet process (HDP) \cite{teh2006hierarchical}. HDP have been successfully applied to many applications involving grouped data, for example, modeling topics within documents comprised of words. For BPA, each pattern is analogous to a document and each bloodstain (ellipse) is analogous to a word. HDP allows for the analysis of multiple sets of data (patterns) by putting a Dirichlet process prior on the base measure.

Consider a number of bloodstain patterns $\bm{p}_1,...,\bm{p}_N$ that share the same bloodletting mechanism $M$ (e.g., impact), where each pattern $\bm{p}_j = (\bm{e}_{j1},...,\bm{e}_{jn_j})$ is represented by a number of ellipses. Assuming the ellipse quintuple follows the SPN distribution, then the HDP model can be expressed by the following formulas:
\begin{align} \label{HDPMM}
\begin{split}
    \bm{e}_{ji}  &\sim \mathcal{SPN}(\Tilde{\bm{\mu}}_{ji},\Tilde{\bm{\Sigma}}_{ji}) \\
    \Tilde{\bm{\mu}}_{ji},\Tilde{\bm{\Sigma}}_{ji} &\sim G_j \\
    G_j &\sim DP(\alpha_M,G_M) \\
    G_M &\sim DP(\alpha_0,G_0) \\
    \alpha_M &\sim Gamma(a, b)
\end{split}
\end{align}
From a generative point of view, the discrete measure $G_j$ dominates the distribution $f_{\bm{p}_j}(\bm{e})$ that generates the ellipses in bloodstain pattern $\bm{p}_j$, and thus $G_j$ can be viewed as an abstraction of pattern $\bm{p}_j$. Moreover, the measure $G_M$ and concentration parameter $\alpha_M$ dominate the distribution of all $G_j$'s, so they can be viewed as an abstraction of the bloodletting mechanism $M$. The model in (\ref{HDPMM}) can be implemented by simply converting Algorithm \ref{algorithm} to the HDP sampling algorithm given in \cite{teh2006hierarchical} (details are provided in Appendix \ref{AppC}). If we train the model (\ref{HDPMM}) with representative bloodstain patterns caused by mechanism $M$, then the likelihood of a new pattern $\bm{p} = (\bm{e}_1,...,\bm{e}_n)$ under the hypothesis $H_M$ that it is caused by $M$ can be estimated by the following 
\begin{align} \label{HDPDE}
\begin{split}
    f(\bm{p}|H_M) = f(\bm{p}|\Hat{\alpha}_M,\Hat{G}_M) &= \int f(\bm{p}|G)dDP(G|\Hat{\alpha}_M,\Hat{G}_M) \\
    &= \int \Bigl\{\prod_{i=1}^{n}  \int \mathcal{SPN}(\bm{e}_j|\Tilde{\bm{\mu}},\Tilde{\bm{\Sigma}})dG(\Tilde{\bm{\mu}},\Tilde{\bm{\Sigma}}) \Bigl\} dDP(G|\Hat{\alpha}_M,\Hat{G}_M)
\end{split}
\end{align}
where $\Hat{\alpha}_M$ and $\Hat{G}_M$ are the posterior mean estimates of $\alpha_M$ and $G_M$ conditional on $\bm{p}_1,...,\bm{p}_N$, and $DP(\cdot|\alpha,G)$ denotes a Dirichlet process measure. The evaluation of the marginal likelihood in (\ref{HDPDE}) including the integral over a Dirichlet process is not straightforward. However, the fact that $G_M$ is sampled from a Dirichlet process and thus is a discrete distribution makes it possible to estimate the marginal likelihood. Details of evaluating (\ref{HDPDE}) are provided in Appendix \ref{AppC}. It is worth noting that \textit{Basu and Chib} \cite{basu2003marginal} proposed a sequential importance sampling method to estimate the marginal likelihood of the data in a DPMM, where the base measure can be a continuous distribution.

\begin{figure}[t]
  \centering
  \resizebox{\textwidth}{!}{%
    \includegraphics{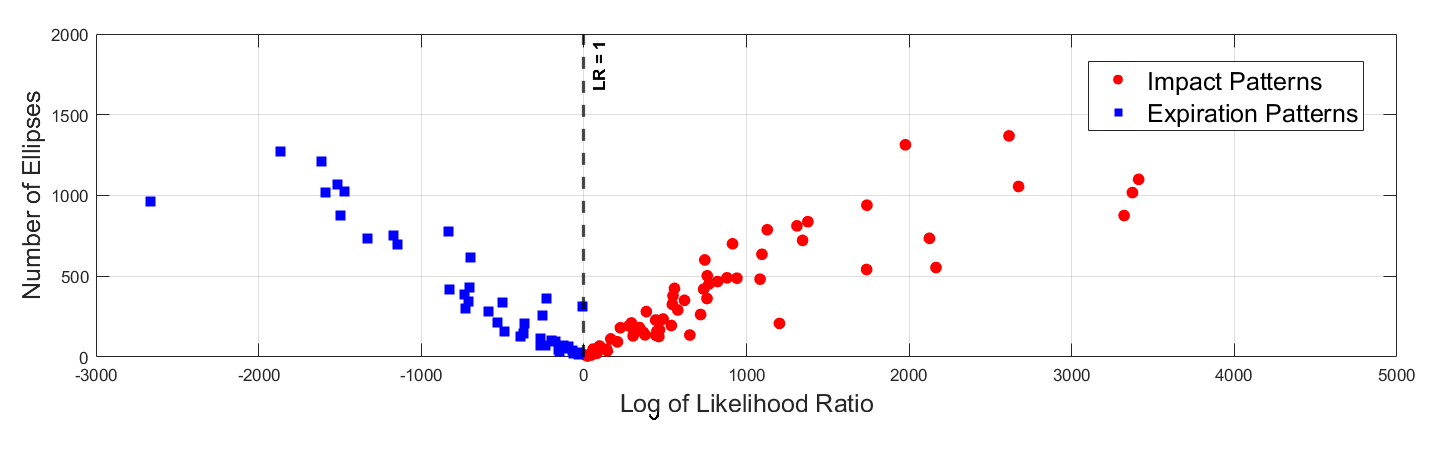}
    }
  \caption{Scatter plot of the log LR and number of ellipses for impact patterns (blue) and expiration patterns (red).}
  \label{LRplot}
\end{figure}

We fit separate HDP models for the two mechanisms for which we have data using 60\% of the bloodstain patterns from each set  (103 impact patterns and 69 expiration patterns). A gamma prior is put on the concentration parameter $\alpha_M$ with hyperparameters $a=b=1$. The rest of the 69 impact patterns and 43 expiration patterns are used to test the performance of the model. For each pattern, its likelihoods under $H_1$ and $H_2$ are calculated via (\ref{HDPDE}) and are used to compute the likelihood ratio defined in (\ref{LRdef}). Figure \ref{LRplot} shows the results; the LR for each test pattern is plotted along with the number of ellipses extracted from that pattern. The LRs are greater than one for all impact patterns and less than one for all expiration patterns. If we use one as threshold for classifying patterns based on the LR, then all test patterns are correctly classified. The magnitude of the log LR (the strength of the evidence) is strongly correlated with the number of ellipses (thus also to the number of bloodstains) because the likelihood of a pattern defined in (\ref{HDPDE}) involves the product of the likelihoods of all ellipses. The correlation seems intuitive in that the LR as a measure of strength of evidence is related to the amount of information that the evidence provides. However, the magnitude of the log LR obtained from many of the patterns is much larger than what might be expected given the uncertainty associated with decisions made by BPA analysts \cite{hicklin2021accuracy}. This likely stems from the lack of diverse and representative impact patterns and expiration patterns. All bloodstain patterns were created in the laboratory with only a few conditions varied, so the model might be learning attributes that are irrelevant to the mechanism. Future study can focus on model calibration and collecting more data to produce more easily interpreted LRs.

\section{Discussion}
\label{Discussion}

In this work we proposed a highly flexible Bayesian nonparametric model to characterize the dependence between linear variables and a directional variable with arbitrary dimension. The multivariate normal distribution was transformed to fit directional-linear data by projecting a number of its dimensions into a unit hypersphere. Then a Dirichlet process mixture model incorporating the semi-projected normal was designed to account for more complex data distributions. A conjugate prior was proposed based on the conditional inverse-Wishart to resolve the identifiability issue raised by the projected normal. Both the clustering and density estimation perspectives were exploited in our experiments.

Future work can focus on more efficient algorithms for posterior inference such as variational methods \cite{blei2006variational, wang2011online}. Another possible direction to extend our approach is to consider modeling the joint distribution of multiple directional variables and linear variables. This requires a more flexible structure for the covariance matrix of the augmented MVN to make the model identifiable, and will also involve developing an appropriate prior distribution.

\begin{appendices}

\section{Sampling the radius \texorpdfstring{$r$}{TEXT}}
\label{appA}

The full conditional distribution of $r$ given in (\ref{sampleR}) has the following form:
\begin{align}\label{rmarg}
f(r) \ \propto \ r^{p-1}\exp{\Bigl\{-\frac{1}{2}Q_3^*(r-\frac{Q_2^*}{Q_3^*})^2\Bigl\}}
\end{align}
\textit{Hernandez-Stumpfhauser et al.} \cite{hernandez2017general} proposed a method to sample $r$ by introducing a latent variable $v$ that has joint density with $r$ given by:
\begin{align}\label{vrjoint}
f(r,v) \ \propto \ r^{p-1}I_{\Bigl(0,\exp{\Bigl\{-\frac{1}{2}Q_3^*(r-\frac{Q_2^*}{Q_3^*})^2\Bigl\}}\Bigl)}(v)I_{(0,\infty)}(r)
\end{align}
Integrating (\ref{vrjoint}) with respect to $v$ yields the marginal distribution of $r$ in (\ref{rmarg}). We can derive the conditional distribution of $v$ and $r$ from (\ref{vrjoint}) and conduct Gibbs sampling.

The conditional distribution of $v$ given $r$ is a uniform distribution:
\begin{align} \label{v_r}
v|r \sim \mathcal{U}\Bigl(0,\exp{\Bigl\{-\frac{1}{2}Q_3^*(r-\frac{Q_2^*}{Q_3^*})^2\Bigl\}}\Bigl)
\end{align}
And the conditional distribution of $r$ given $v$ is:
\begin{align*}
f(r|v) \ \propto \ r^{p-1}I_{\Bigl(\frac{Q_2^*}{Q_3^*}+max\Bigl\{-\frac{Q_2^*}{Q_3^*},-\sqrt{\frac{-2\ln{v}}{Q_3^*}}\Bigl\},\frac{Q_2^*}{Q_3^*}+\sqrt{\frac{-2\ln{v}}{Q_3^*}}\Bigl)}(r)
\end{align*}
By using the inverse cumulative distribution function technique we get
\begin{gather} \label{computeR}
r=[(\eta^p_2-\eta^p_1)w+\eta^p_1]^\frac{1}{p}
\end{gather}
where
\begin{gather*}
    w \sim \mathcal{U}(0,1), \quad \eta_1 = \frac{Q_2^*}{Q_3^*}+max\Bigl\{-\frac{Q_2^*}{Q_3^*},-\sqrt{\frac{-2\ln{v}}{Q_3^*}}\Bigl\}, \quad \eta_2 = \frac{Q_2^*}{Q_3^*}+\sqrt{\frac{-2\ln{v}}{Q_3^*}}
\end{gather*}
One issue with this method is that when the difference between $r$ and $\frac{Q_2^*}{Q_3^*}$ is large, underflow of $v$ may occur and lead to overflow of $r$. We suggest directly sample $\ln{v}$ using the inverse cumulative distribution function technique to avoid this issue. Instead of sampling $v$ from (\ref{v_r}), sample $s \sim \mathcal{U}(0,1)$, compute $\ln{v}$:
\begin{gather*}
    \ln{v}=\ln{s}-\frac{1}{2}Q_3^*(r-\frac{Q_2^*}{Q_3^*})^2
\end{gather*}
and compute $r$ via (\ref{computeR}).

\section{Properties of the inverse-Wishart distribution}
\label{appB}

To prove that the partitioned inverse-Wishart distribution has the properties given in (\ref{CIWprop}), let matrix $\bm{\Gamma}$ follow the Wishart distribution $\mathcal{W}(\bm{R},\nu)$ and partition $\bm{\Gamma}$ and $\bm{R}$ as:
\begin{align*}
    \bm{\Gamma} = \begin{pmatrix}
    \bm{\Gamma}_{11} & \bm{\Gamma}_{12} \\
    \bm{\Gamma}_{21} & \bm{\Gamma}_{22}
    \end{pmatrix},\quad
    \bm{R} = \begin{pmatrix}
    \bm{R}_{11} & \bm{R}_{12} \\
    \bm{R}_{21} & \bm{R}_{22}
    \end{pmatrix}
\end{align*}
Here $\bm{\Gamma}$ and $\bm{R}$ are $d\times d$ matrices, and $\bm{\Gamma}_{ij}$ and $\bm{R}_{ij}$ are $d_i\times d_j$ matrices ($d_1+d_2=d$). Denote $\bm{\Gamma}_{11\cdot2}=\bm{\Gamma}_{11}-\bm{\Gamma}_{12}\bm{\Gamma}_{22}^{-1}\bm{\Gamma}_{21}$ and $\bm{R}_{11\cdot2}=\bm{R}_{11}-\bm{R}_{12}\bm{R}_{22}^{-1}\bm{R}_{21}$, then according to \textbf{Theorem 3.3.9} in \cite{gupta1999matrix}:
\begin{align*}
    &\text{(i) }\bm{\Gamma}_{22}\sim\mathcal{W}(\bm{R}_{22}, \nu)  \\
    &\text{(ii) }\bm{\Gamma}_{11\cdot2} \sim \mathcal{W}(\bm{R}_{11\cdot2}, \nu - d_2) \\
    &\text{(iii) }\bm{\Gamma}_{11\cdot2} \text{ and } (\bm{\Gamma}_{12}, \bm{\Gamma}_{22}) \text{ are independent} \\
    &\text{(iv) }\bm{vec}(\bm{\Gamma}_{12}) | \bm{\Gamma}_{22} \sim \mathcal{N}_{d_1\times d_2}[\bm{vec}(\bm{R}_{12}\bm{R}_{22}^{-1}\bm{\Gamma}_{22}), \bm{\Gamma}_{22} \otimes \bm{R}_{11\cdot2}]
\end{align*}
Property (iv) is slightly different from the version in the book, because here we directly use $\bm{vec}(\cdot)$ on the matrix to denote the matrix-variate normal distribution while the book uses the vectorization of the transpose of the matrix.

Let $\Tilde{\bm{\Sigma}}=\bm{\Gamma}^{-1}$ and $\bm{S}=\bm{R}^{-1}$, then by definition $\Tilde{\bm{\Sigma}} \sim \mathcal{IW}(\bm{S}, \nu)$. According to (\ref{partitionS}) and the properties of the inverse of a partitioned matrix we have the following relations:
\begin{align*}
    \bm{\Sigma}_{11} &= \bm{\Gamma}_{11\cdot2}^{-1} \\ \bm{S}_{11} &= \bm{R}_{11\cdot2}^{-1} \\
    \bm{\Sigma}_{22\cdot1} &= \bm{\Gamma}_{22}^{-1}\\
    \bm{S}_{22\cdot1} &= \bm{R}_{22}^{-1} \\
    \bm{\Sigma}_{11}^{-1}\bm{\Sigma}_{12} &= -\bm{\Gamma}_{12}\bm{\Gamma}_{22}^{-1} \\
    \bm{S}_{11}^{-1}\bm{S}_{12} &= -\bm{R}_{12}\bm{R}_{22}^{-1} 
\end{align*}
Considering the first four equations, properties (a) and (d) in (\ref{CIWprop}) are equivalent to properties (ii) and (i) above.

Because $\bm{\Sigma}_{11}$ can be derived from $\bm{\Gamma}_{11\cdot2}$, and $(\bm{\Sigma}_{11}^{-1}\bm{\Sigma}_{12}, \bm{\Sigma}_{22\cdot1})$ can be derived from $(\bm{\Gamma}_{12}, \bm{\Gamma}_{22})$, then according to property (iii), $\bm{\Sigma}_{11}$ is independent of $(\bm{\Sigma}_{11}^{-1}\bm{\Sigma}_{12}, \bm{\Sigma}_{22\cdot1})$. Hence, property (b) is true.

From the relations above  we can rewrite property (iv) in terms of $\bm{\Sigma}_{ij}$ and $\bm{S}_{ij}$ as
\begin{gather*}
    \bm{vec}(-\bm{\Sigma}_{11}^{-1}\bm{\Sigma}_{12}\bm{\Sigma}_{22\cdot1}^{-1}) | \bm{\Sigma}_{22\cdot1} \sim \mathcal{N}_{d_1\times d_2}[\bm{vec}(-\bm{S}_{11}^{-1}\bm{S}_{12}\bm{\Sigma}_{22\cdot1}^{-1}), \bm{\Sigma}_{22\cdot1}^{-1} \otimes \bm{S}_{11}^{-1}]
\end{gather*}
Then according to \textbf{Theorem 2.3.10} in \cite{gupta1999matrix} (again the notation is slightly different due to vectorization), property (c) can be derived by right multiplying by $-\bm{\Sigma}_{22\cdot1}$.

\section{Implementation of the hierarchical DPSPN}
\label{AppC}

Our goal of fitting a hierarchical DPSPN is to obtain an estimate of the measure $G_M$ and the concentration parameter $\alpha_M$ in (\ref{HDPMM}) that characterize the bloodstain pattern generation mechanism $M$. Then we can use them in (\ref{HDPDE}) to evaluate the likelihood of a new pattern assuming it is generated by mechanism $M$. \textit{Teh et al.} \cite{teh2006hierarchical} proposed an algorithm by direct assignment that can be applied to sample $G_M$ from the posterior distribution by expressing $G_M$ based on the stick-breaking representation:
\begin{gather}
    G_M = \sum_{k=1}^K \beta_k\delta_{\bm{\varphi}_k} + \beta_uG_u
\end{gather}
Here $\{\beta_k\}_{k=1}^K$ are the mixture probabilities and $\bm{\varphi}_k=(\Tilde{\bm{\mu}}_k,\Tilde{\bm{\Sigma}}_k)$ are the mixture parameters where $\delta_{\bm{\varphi}_k}$ denotes a Dirac delta distribution at $\bm{\varphi}_k$. $\beta_u$ is the probability of creating a new cluster and $G_u$ is a measure sampled from $DP(\alpha_0,G_0)$. The algorithm starts by sampling the clustering assignment for each observation as follows:
\begin{align} \label{samplecji}
    P(c_{ji}=k|c_{-ji},\bm{z}_{ji})\ \propto
    \begin{cases}
    (n_{-i,k}^j+\alpha_M\beta_k) f(\bm{z}_{ji}|\bm{\Psi}^k)\qquad &\text{if }c\text{ represents an existing cluster}  \\
    \alpha_M\beta_uf(\bm{z}_{ji}|\bm{\Psi}^0) \qquad &\text{if }c\text{ represents a new cluster}
    \end{cases}
\end{align}
where $n_{-i,k}^j$ is the number of ellipses assigned to cluster $k$ in pattern $j$ excluding ellipse $i$. The likelihood function $f(\bm{z}|\bm{\Psi})$ can be evaluated via (\ref{datamarginal}). Next, an intermediate variable $m_{jk}$ is sampled based on the following distribution:
\begin{align} \label{samplemjk}
    P(m_{jk}=m|\bm{c},\bm{\beta})\ \propto\ s(n_k^j,m)(\alpha_M\beta_k)^m, \quad m=1,...,n_k^j
\end{align}
where $n_k^j$ is the number of ellipses assigned to cluster $k$ in pattern $j$ and $s(n,m)$ is the unsigned Stirling number of the first kind. From the perspective of the Chinese restaurant process \cite{aldous1985exchangeability}, $m_{jk}$ denotes the number of tables assigned to cluster $k$ in restaurant $j$, and its conditional distribution given in (\ref{samplemjk}) was proved by \textit{Antoniak} \cite{antoniak1974mixtures}. Finally, we can sample $(\beta_1,...,\beta_K,\beta_u)$ from a Dirichlet distribution
\begin{align} \label{samplebeta}
    (\beta_1,...,\beta_K,\beta_u)\sim Dir(m_{\cdot 1},...,m_{\cdot K},\alpha_0)
\end{align}
where $m_{\cdot k} = \sum_{j=1}^Jm_{jk}$ is the number of tables assigned to cluster $k$. Algorithm \ref{algorithm2} provides the pseudocode to sample from the hierarchical DPSPN.

\begin{algorithm} 
\caption{Gibbs Sampler for the hierarchical DPSPN}\label{algorithm2}
\begin{algorithmic}
\State Random initialization of $\{c_{ji}\}_{i=1,j=1}^{n_j,J}$ and $\{r_{ji}\}_{i=1,j=1}^{n_j,J}$
\State $K = \#$ of clusters 
\For{$iter = 1$ to $M$}
    \State update $\{x_{ji}\}_{i=1,j=1}^{n_j,J}$ with $\{r_{ji}\}_{i=1,j=1}^{n_j,J}$ using (\ref{car2sph}) for $j=1,...,J$
    \For{$j = 1$ to $N$ and $i = 1$ to $n_j$}
        \State remove $z_{ji}$ from its current cluster $c_{ji}$
        \State update the posterior parameter $\bm{\Psi}$ of cluster $c_{ji}$ using (\ref{postpara})
        \State if the cluster is empty, remove it and decrease $K$
        \State sample a new value for $c_{ji}$ from $P(c_{ji}|c_{-ji},\bm{z}_{ji})$ according to (\ref{samplecji})
        \State add $\bm{z_{ji}}$ to cluster $c_{ji}$
        \State update $\bm{\Psi}^{c_{ji}}$ using (\ref{postpara})
        \State if a new cluster is created, increase $K$
    \EndFor
    
    \For{$j = 1$ to $J$ and $k = 1$ to $K$}
        \State sample $m_{jk}$ according to (\ref{samplemjk})
    \EndFor
    \State sample $(\beta_1,...,\beta_K,\beta_u)$ according to (\ref{samplebeta})
    \For{$k = 1$ to $K$}
        \State sample $\bm{\varphi}_k=(\Tilde{\bm{\mu}}^k,\Tilde{\bm{\Sigma}}^k)$ from $\mathcal{NCIW}(\bm{\Psi}^k)$
    \EndFor
    \For{$j = 1$ to $J$ and $i = 1$ to $n_j$}
        \State sample $r_{ji}$ from $f(r|\bm{\theta}_{ji},\bm{y}_{ji},\Tilde{\bm{\mu}}^{c_{ji}},\Tilde{\bm{\Sigma}}^{c_{ji}})$ using (\ref{sampleR})
    \EndFor
    \State update the concentration parameter $\alpha_M$ (optional, see \cite{teh2006hierarchical})
\EndFor
\end{algorithmic}
\end{algorithm}

Formula (\ref{HDPDE}) computes the likelihood of a pattern and its evaluation involves estimating $G_M$. From the Gibbs sampling algorithm we can sample $\{\beta_k\}_{k=1}^{K}$ and $\{\bm{\varphi}_k\}_{k=1}^{K}$. In the application to the BPA data, as the number of clusters increases through iterations, $\beta_u$ becomes significantly smaller than one. As a result, and for computational convenience, we approximate $G_M$ by cutting off the term $\beta_uG_u$ as follows:
\begin{align}
    \Hat{G}_M = \sum_{k=1}^K \Hat{\beta}_k\delta_{\Hat{\bm{\varphi}}_k} \quad \text{where} \quad \Hat{\beta}_k = \frac{\beta_k}{\sum_{k'=1}^K \beta_{k'}}
\end{align}
Let $G$ be sampled from $DP(\Hat{\alpha}_M, \Hat{G}_M)$. Since $\Hat{G}_M$ is a finite discrete distribution, so is $G$:
\begin{align}
    G = \sum_{k=1}^K \pi_k\delta_{\Hat{\bm{\varphi}}_k}
\end{align}
The probability weights $\{\pi_k\}_{k=1}^{K}$ can be shown to follow a Dirichlet distribution using the derivation in \cite{teh2006hierarchical}:
\begin{align} \label{dirdist}
    (\pi_1,...,\pi_K)\sim Dir(\Hat{\alpha}_M\Hat{\beta}_1,...,\Hat{\alpha}_M\Hat{\beta}_K)
\end{align}
Now we can rewrite (\ref{HDPDE}) in terms of $\{\pi_k\}_{k=1}^{K}$:
\begin{align}
f(\bm{p}|\Hat{\alpha}_M,\Hat{G}_M) &= \int \Bigl\{\prod_{i=1}^{n}  \int \mathcal{SPN}(\bm{e}_j|\Tilde{\bm{\mu}},\Tilde{\bm{\Sigma}})dG(\Tilde{\bm{\mu}},\Tilde{\bm{\Sigma}}) \Bigl\} dDP(G|\Hat{\alpha}_M,\Hat{G}_M) \\
&= \int \Bigl\{\prod_{i=1}^{n}  \sum_{k=1}^K \pi_k\mathcal{SPN}(\bm{e}_j|\Tilde{\bm{\mu}}_k,\Tilde{\bm{\Sigma}}_k) \Bigl\}p(\pi_1,...,\pi_K)d\pi_1...d\pi_K \label{mclkhd}
\end{align}
Analytical evaluation of (\ref{mclkhd}) is possible when $K$ and $n$ is small. An alternative way to estimate the integral is to use the Monte Carlo approach by sampling $\{\pi_k\}_{k=1}^{K}$ from (\ref{dirdist}).

\end{appendices}

\section*{Acknowledgments}
This work was funded by the Center for Statistics and Applications in Forensic Evidence (CSAFE) through Cooperative Agreements 70NANB15H176 and 70NANB20H019 between NIST and Iowa State University, which includes activities carried out at Carnegie Mellon University, Duke University, University of California Irvine, University of Virginia, West Virginia University, University of Pennsylvania, Swarthmore College and University of Nebraska, Lincoln. The authors would like to thank Tianyu Pan for his discussion on the model development and Ziyi Song for his great advice that leads to the use of SALSO algorithm. The authors are also grateful to the late Michael Taylor for providing the data and for numerous conversations that impacted the work.

\bibliographystyle{unsrt}  
\bibliography{references}

\begin{thebibliography}{10}

\bibitem{jammalamadaka2006effect}
S~R Jammalamadaka and U~J Lund.
\newblock The effect of wind direction on ozone levels: a case study.
\newblock {\em Environmental and Ecological Statistics}, 13(3):287--298, 2006.

\bibitem{nunez2015bayesian}
G~Nu{\~n}ez-Antonio, M~C Aus{\'\i}n, and M~P Wiper.
\newblock Bayesian nonparametric models of circular variables based on
  {D}irichlet process mixtures of normal distributions.
\newblock {\em Journal of Agricultural, Biological, and Environmental
  Statistics}, 20(1):47--64, 2015.

\bibitem{kucner2017enabling}
T~P Kucner, M~Magnusson, E~Schaffernicht, V~H Bennetts, and A~J Lilienthal.
\newblock Enabling flow awareness for mobile robots in partially observable
  environments.
\newblock {\em IEEE Robotics and Automation Letters}, 2(2):1093--1100, 2017.

\bibitem{palmieri2017kinodynamic}
L~Palmieri, T~P Kucner, M~Magnusson, A~J Lilienthal, and K~O Arras.
\newblock Kinodynamic motion planning on {G}aussian mixture fields.
\newblock In {\em 2017 IEEE International Conference on Robotics and Automation
  (ICRA)}, pages 6176--6181. IEEE, 2017.

\bibitem{hasnat2014unsupervised}
M~A Hasnat, O~Alata, and A~Tr{\'e}meau.
\newblock Unsupervised clustering of depth images using {W}atson mixture model.
\newblock In {\em 2014 22nd International Conference on Pattern Recognition},
  pages 214--219. IEEE, 2014.

\bibitem{roy2016swgmm}
A~Roy, S~K Parui, and U~Roy.
\newblock Swgmm: a semi-wrapped {G}aussian mixture model for clustering of
  circular--linear data.
\newblock {\em Pattern Analysis and Applications}, 19(3):631--645, 2016.

\bibitem{watson1982distributions}
G~S Watson.
\newblock Distributions on the circle and sphere.
\newblock {\em Journal of Applied Probability}, 19(A):265--280, 1982.

\bibitem{mardia1975statistics}
K~V Mardia.
\newblock Statistics of directional data.
\newblock {\em Journal of the Royal Statistical Society: Series B
  (Methodological)}, 37(3):349--371, 1975.

\bibitem{wang2013directional}
F~Wang and A~E Gelfand.
\newblock Directional data analysis under the general projected normal
  distribution.
\newblock {\em Statistical Methodology}, 10(1):113--127, 2013.

\bibitem{hernandez2017general}
D~Hernandez-Stumpfhauser, F~J Breidt, and M~J van~der Woerd.
\newblock The general projected normal distribution of arbitrary dimension:
  modeling and {B}ayesian inference.
\newblock {\em Bayesian Analysis}, 12(1):113--133, 2017.

\bibitem{collett1981discriminating}
D~Collett and T~Lewis.
\newblock Discriminating between the von {M}ises and wrapped normal
  distributions.
\newblock {\em Australian Journal of Statistics}, 23(1):73--79, 1981.

\bibitem{mardia2000directional}
K~V Mardia, P~E Jupp, and K~V Mardia.
\newblock {\em Directional Statistics}, volume~2.
\newblock Wiley Online Library, 2000.

\bibitem{pewsey2021recent}
A~Pewsey and E~Garc{\'\i}a-Portugu{\'e}s.
\newblock Recent advances in directional statistics.
\newblock {\em Test}, 30(1):1--58, 2021.

\bibitem{wang2014modeling}
F~Wang and A~E Gelfand.
\newblock Modeling space and space-time directional data using projected
  {G}aussian processes.
\newblock {\em Journal of the American Statistical Association},
  109(508):1565--1580, 2014.

\bibitem{rodriguez2020bayesian}
C~E Rodr{\'\i}guez, G~N{\'u}{\~n}ez-Antonio, and G~Escarela.
\newblock A {B}ayesian mixture model for clustering circular data.
\newblock {\em Computational Statistics \& Data Analysis}, 143:106842, 2020.

\bibitem{carta2008statistical}
J~A Carta, C~Bueno, and P~Ram{\'\i}rez.
\newblock Statistical modelling of directional wind speeds using mixtures of
  von {M}ises distributions: case study.
\newblock {\em Energy Conversion and Management}, 49(5):897--907, 2008.

\bibitem{fernandez2004circular}
J~J Fern{\'a}ndez-Dur{\'a}n.
\newblock Circular distributions based on nonnegative trigonometric sums.
\newblock {\em Biometrics}, 60(2):499--503, 2004.

\bibitem{fernandez2007models}
J~J Fern{\'a}ndez-Dur{\'a}n.
\newblock Models for circular--linear and circular--circular data constructed
  from circular distributions based on nonnegative trigonometric sums.
\newblock {\em Biometrics}, 63(2):579--585, 2007.

\bibitem{johnson1978some}
R~A Johnson and T~E Wehrly.
\newblock Some angular-linear distributions and related regression models.
\newblock {\em Journal of the American Statistical Association},
  73(363):602--606, 1978.

\bibitem{carta2008joint}
J~A Carta, P~Ramirez, and C~Bueno.
\newblock A joint probability density function of wind speed and direction for
  wind energy analysis.
\newblock {\em Energy Conversion and Management}, 49(6):1309--1320, 2008.

\bibitem{soukissian2014probabilistic}
T~H Soukissian.
\newblock Probabilistic modeling of directional and linear characteristics of
  wind and sea states.
\newblock {\em Ocean Engineering}, 91:91--110, 2014.

\bibitem{zhang2018investigation}
L~Zhang, Q~Li, Y~Guo, Z~Yang, and L~Zhang.
\newblock An investigation of wind direction and speed in a featured wind farm
  using joint probability distribution methods.
\newblock {\em Sustainability}, 10(12):4338, 2018.

\bibitem{roy2017jclmm}
A~Roy, A~Pal, and U~Garain.
\newblock {JCLMM}: A finite mixture model for clustering of circular-linear
  data and its application to psoriatic plaque segmentation.
\newblock {\em Pattern Recognition}, 66:160--173, 2017.

\bibitem{mastrantonio2018joint}
G~Mastrantonio.
\newblock The joint projected normal and skew-normal: A distribution for
  poly-cylindrical data.
\newblock {\em Journal of Multivariate Analysis}, 165:14--26, 2018.

\bibitem{pukkila1988pattern}
T~M Pukkila and C~R Rao.
\newblock Pattern recognition based on scale invariant discriminant functions.
\newblock {\em Information Sciences}, 45(3):379--389, 1988.

\bibitem{ferguson1973bayesian}
T~S Ferguson.
\newblock A {B}ayesian analysis of some nonparametric problems.
\newblock {\em The Annals of Statistics}, pages 209--230, 1973.

\bibitem{escobar1995bayesian}
M~D Escobar and M~West.
\newblock Bayesian density estimation and inference using mixtures.
\newblock {\em Journal of the American Statistical Association},
  90(430):577--588, 1995.

\bibitem{teh2006hierarchical}
Y~W Teh, M~I Jordan, M~J Beal, and D~M Blei.
\newblock Hierarchical {D}irichlet processes.
\newblock {\em Journal of the American Statistical Association},
  101(476):1566--1581, 2006.

\bibitem{neal2000markov}
R~M Neal.
\newblock Markov chain sampling methods for {D}irichlet process mixture models.
\newblock {\em Journal of Computational and Graphical Statistics},
  9(2):249--265, 2000.

\bibitem{ishwaran2002exact}
H~Ishwaran and M~Zarepour.
\newblock Exact and approximate sum representations for the {D}irichlet
  process.
\newblock {\em Canadian Journal of Statistics}, 30(2):269--283, 2002.

\bibitem{maceachern1994estimating}
S~N MacEachern.
\newblock Estimating normal means with a conjugate style {D}irichlet process
  prior.
\newblock {\em Communications in Statistics-Simulation and Computation},
  23(3):727--741, 1994.

\bibitem{Märtens2018}
K~Märtens.
\newblock Mixture{M}odels.
\newblock \url{https://github.com/kasparmartens/mixtureModels}, 2018.

\bibitem{gelman1992inference}
A~Gelman and D~B Rubin.
\newblock Inference from iterative simulation using multiple sequences.
\newblock {\em Statistical Science}, pages 457--472, 1992.

\bibitem{brooks1998general}
S~P Brooks and A~Gelman.
\newblock General methods for monitoring convergence of iterative simulations.
\newblock {\em Journal of Computational and Graphical Statistics},
  7(4):434--455, 1998.

\bibitem{dahl2022search}
D~B Dahl, D~J Johnson, and P~M{\"u}ller.
\newblock Search algorithms and loss functions for {B}ayesian clustering.
\newblock {\em Journal of Computational and Graphical Statistics}, pages 1--13,
  2022.

\bibitem{meilua2007comparing}
M~Meil{\u{a}}.
\newblock Comparing clusterings—an information based distance.
\newblock {\em Journal of Multivariate Analysis}, 98(5):873--895, 2007.

\bibitem{hubert1985comparing}
L~Hubert and P~Arabie.
\newblock Comparing partitions.
\newblock {\em Journal of Classification}, 2(1):193--218, 1985.

\bibitem{rand1971objective}
W~M Rand.
\newblock Objective criteria for the evaluation of clustering methods.
\newblock {\em Journal of the American Statistical Association},
  66(336):846--850, 1971.

\bibitem{xuan2010information}
N~X Vinh, J~Epps, and J~Bailey.
\newblock Information theoretic measures for clusterings comparison: variants,
  properties, normalization and correction for chance.
\newblock {\em Journal of Machine Learning Research}, 11:2837–2854, 2010.

\bibitem{wang2007object}
L~Wang, J~Shi, G~Song, and I~Shen.
\newblock Object detection combining recognition and segmentation.
\newblock In {\em Asian Conference on Computer Vision}, pages 189--199.
  Springer, 2007.

\bibitem{pham2000survey}
D~L Pham, C~Xu, and J~L Prince.
\newblock A survey of current methods in medical image segmentation.
\newblock {\em Annual Review of Biomedical Engineering}, 2(3):315--337, 2000.

\bibitem{kato2006markov}
Z~Kato and T~Pong.
\newblock A {M}arkov random field image segmentation model for color textured
  images.
\newblock {\em Image and Vision Computing}, 24(10):1103--1114, 2006.

\bibitem{shafarenko1997automatic}
L~Shafarenko, M~Petrou, and J~Kittler.
\newblock Automatic watershed segmentation of randomly textured color images.
\newblock {\em IEEE Transactions on Image Processing}, 6(11):1530--1544, 1997.

\bibitem{mignotte2008segmentation}
M~Mignotte.
\newblock Segmentation by fusion of histogram-based $k$-means clusters in
  different color spaces.
\newblock {\em IEEE Transactions on Image Processing}, 17(5):780--787, 2008.

\bibitem{ihaka2003colour}
R~Ihaka.
\newblock Colour for presentation graphics.
\newblock In {\em Proceedings of DSC}, volume~2, 2003.

\bibitem{martin2001database}
D~Martin, C~Fowlkes, D~Tal, and J~Malik.
\newblock A database of human segmented natural images and its application to
  evaluating segmentation algorithms and measuring ecological statistics.
\newblock In {\em Proceedings Eighth IEEE International Conference on Computer
  Vision. ICCV 2001}, volume~2, pages 416--423. IEEE, 2001.

\bibitem{unnikrishnan2005measure}
R~Unnikrishnan, C~Pantofaru, and M~Hebert.
\newblock A measure for objective evaluation of image segmentation algorithms.
\newblock In {\em 2005 IEEE Computer Society Conference on Computer Vision and
  Pattern Recognition (CVPR'05)-Workshops}, pages 34--34. IEEE, 2005.

\bibitem{freixenet2002yet}
J~Freixenet, X~Munoz, D~Raba, J~Mart{\'\i}, and X~Cuf{\'\i}.
\newblock Yet another survey on image segmentation: region and boundary
  information integration.
\newblock In {\em European Conference on Computer Vision}, pages 408--422.
  Springer, 2002.

\bibitem{yang2008unsupervised}
A~Y Yang, J~Wright, Y~Ma, and S~S Sastry.
\newblock Unsupervised segmentation of natural images via lossy data
  compression.
\newblock {\em Computer Vision and Image Understanding}, 110(2):212--225, 2008.

\bibitem{stephens2000dealing}
M~Stephens.
\newblock Dealing with label switching in mixture models.
\newblock {\em Journal of the Royal Statistical Society: Series B (Statistical
  Methodology)}, 62(4):795--809, 2000.

\bibitem{roy2007beta}
A~Roy, S~K Parui, and U~Roy.
\newblock A beta mixture model based approach to text extraction from color
  images.
\newblock In {\em Advances in Pattern Recognition}, pages 321--326. World
  Scientific, 2007.

\bibitem{roy2012mixture}
A~Roy, S~K Parui, and U~Roy.
\newblock A mixture model of circular-linear distributions for color image
  segmentation.
\newblock {\em International Journal of Computer Applications}, 58(9), 2012.

\bibitem{boutemedjet2008hybrid}
S~Boutemedjet, N~Bouguila, and D~Ziou.
\newblock A hybrid feature extraction selection approach for high-dimensional
  non-{G}aussian data clustering.
\newblock {\em IEEE Transactions on Pattern Analysis and Machine Intelligence},
  31(8):1429--1443, 2008.

\bibitem{damelio2001bloodstain}
R~Damelio and R~M Gardner.
\newblock {\em Bloodstain Pattern Analysis: with an Introduction to Crime Scene
  Reconstruction}.
\newblock CRC press, 2001.

\bibitem{national2009strengthening}
{National Research Council} et~al.
\newblock {\em Strengthening Forensic Science in the United States: A Path
  Forward}.
\newblock National Academies Press, 2009.

\bibitem{hicklin2021accuracy}
R~A Hicklin, K~R Winer, P~E Kish, C~L Parks, W~Chapman, K~Dunagan,
  N~Richetelli, E~G Epstein, M~A Ausdemore, and T~A Busey.
\newblock Accuracy and reproducibility of conclusions by forensic bloodstain
  pattern analysts.
\newblock {\em Forensic Science International}, 325:110856, 2021.

\bibitem{arthur2018automated}
R~M Arthur, J~Hoogenboom, M~Baiker, M~C Taylor, and K~G de~Bruin.
\newblock An automated approach to the classification of impact spatter and
  cast-off bloodstain patterns.
\newblock {\em Forensic Science International}, 289:310--319, 2018.

\bibitem{liu2020automatic}
Y~Liu, D~Attinger, and K~de~Brabanter.
\newblock Automatic classification of bloodstain patterns caused by gunshot and
  blunt impact at various distances.
\newblock {\em Journal of Forensic Sciences}, 65(3):729--743, 2020.

\bibitem{zou2022towards}
T~Zou and H~S Stern.
\newblock Towards a likelihood ratio approach for bloodstain pattern analysis.
\newblock {\em Forensic Science International}, page 111512, 2022.

\bibitem{zou2021recognition}
T~Zou, T~Pan, M~Taylor, and H~S Stern.
\newblock Recognition of overlapping elliptical objects in a binary image.
\newblock {\em Pattern Analysis and Applications}, 24(3):1193--1206, 2021.

\bibitem{basu2003marginal}
S~Basu and S~Chib.
\newblock Marginal likelihood and {B}ayes factors for {D}irichlet process
  mixture models.
\newblock {\em Journal of the American Statistical Association},
  98(461):224--235, 2003.

\bibitem{blei2006variational}
D~M Blei and M~I Jordan.
\newblock Variational inference for {D}irichlet process mixtures.
\newblock {\em Bayesian Analysis}, 1(1):121--143, 2006.

\bibitem{wang2011online}
C~Wang, J~Paisley, and D~M Blei.
\newblock Online variational inference for the hierarchical {D}irichlet
  process.
\newblock In {\em Proceedings of the Fourteenth International Conference on
  Artificial Intelligence and Statistics}, pages 752--760. JMLR Workshop and
  Conference Proceedings, 2011.

\bibitem{gupta1999matrix}
A~K Gupta and D~K Nagar.
\newblock {\em Matrix Variate Distributions}.
\newblock Chapman and Hall/CRC, 1999.

\bibitem{aldous1985exchangeability}
D~J Aldous.
\newblock Exchangeability and related topics.
\newblock In {\em {\'E}cole d'{\'E}t{\'e} de Probabilit{\'e}s de Saint-Flour
  XIII—1983}, pages 1--198. Springer, 1985.

\bibitem{antoniak1974mixtures}
C~E Antoniak.
\newblock Mixtures of {D}irichlet processes with applications to {B}ayesian
  nonparametric problems.
\newblock {\em The Annals of Statistics}, pages 1152--1174, 1974.

\end{thebibliography}

\end{document}